\documentclass[twocolumn,showpacs,preprintnumbers,amsmath,amssymb]{revtex4}
\usepackage{epsfig}
\usepackage{dcolumn}
\usepackage{bm}

\newcommand{\remove}[1]{}
\newcommand{\dd}{\mathrm{d}}

\def\be{\begin{equation}}
\def\ee{\end{equation}}

\newcommand{\beq}{\begin{equation}}
\newcommand{\eeq}{\end{equation}}
\newcommand{\beqa}{\begin{eqnarray}}
\newcommand{\eeqa}{\end{eqnarray}}

\renewcommand{\pl}{\partial}

\newcommand{\tg}{{\tilde{g}}}

\newcommand{\cG}{{\cal G}}

\newcommand{\cM}{{\cal M}}

\newcommand{\rhob}{\overline{\rho}}

\newcommand{\bea}{\begin{array}}
\newcommand{\ea}{\end{array}}

\newcommand{\MPl}{M_{\rm Pl}}

\begin{document}

\title{K-mouflage Cosmology: the Background Evolution}

\author{Philippe Brax}
\affiliation{Institut de Physique Th\'eorique,\\
CEA, IPhT, F-91191 Gif-sur-Yvette, C\'edex, France\\
CNRS, URA 2306, F-91191 Gif-sur-Yvette, C\'edex, France}
\author{Patrick Valageas}
\affiliation{Institut de Physique Th\'eorique,\\
CEA, IPhT, F-91191 Gif-sur-Yvette, C\'edex, France\\
CNRS, URA 2306, F-91191 Gif-sur-Yvette, C\'edex, France}
\vspace{.2 cm}

\date{\today}
\vspace{.2 cm}

\begin{abstract}
We study the cosmology of K-mouflage theories at the background level. We show that the effects of the scalar field
are suppressed at high matter density in the early Universe and only play a role in the late time Universe where the deviations of the Hubble rate from its $\Lambda$-CDM counterpart can be
of the order five percent for redshifts $1 \lesssim z \lesssim 5$. Similarly, we find that the equation of state can cross the phantom divide in the recent past and even diverge when the effective scalar energy density
goes negative and subdominant compared to matter, preserving the positivity of the squared Hubble rate. These features are present in models for which Big Bang Nucleosynthesis is not affected. We analyze the fate of K-mouflage when the nonlinear kinetic terms give rise to
ghosts, particle excitations with negative energy.
In this case, we find that the K-mouflage theories can only be considered as an effective description of the Universe at low energy below $1$ keV. In the safe ghost-free models, we find that the equation of state always diverges in the past and changes significantly by a few percent since $z\lesssim 1$.

\end{abstract}

 \vskip2pc

\maketitle

\section{Introduction}
\label{Introduction}

Scalar fields could be playing a role in late time cosmology and have something to do with the recent acceleration of the expansion of the Universe\cite{Copeland:2006wr}. They could also induce modifications of gravity on large scales \cite{Khoury:2010xi}
which may be within the reach of the forthcoming EUCLID mission \cite{Amendola:2012ys}. For instance  one always expects at least one scalar field to be present in massive extensions of General Relativity (GR) which could lead to such deviations\cite{deRham:2012az}. In both cases, the mass of the scalar field is very small, implying the possibility of the existence of a scalar fifth force.   All in all, scalar fields with low masses acting on very large scales of the Universe are ubiquitous. Nevertheless, such scalars have never
manifested themselves in the Solar System or the laboratory where deviations from General Relativity have been painstakingly sought for\cite{Will:2004nx}. It has been recently advocated that this could be the result of the screening of these scalar fields in dense environments\cite{Brax:2012yi,Brax:2013ida}. As a result, they would be nearly invisible locally while acting as  full fledged modifications of the dynamics of the Universe on large scales.

In this paper, we shall focus on one particular type of screening: K-mouflage\cite{Babichev:2009ee,Brax:2012jr}.
More precisely, we focus on models that have the simplest K-essence form \cite{Armendariz-Picon:1999aa}, where the nonstandard kinetic term is only a nonlinear function of $(\pl\varphi)^2$.
This is a subclass of the possible models that one can build with nonstandard kinetic terms,
that may also depend on the field $\varphi$ or higher derivatives $\pl^2\varphi$ \cite{Babichev:2009ee}.
These models partake in the three types of screening mechanisms which are compatible with second order equations of motion for higher order scalar field theories. In K-mouflage theories, the equations of motion are always second order but the Hamiltonian corresponding to the scalar energy density may be negative for large values of the field's time derivative, depending on the form of the kinetic term. However, even in such cases, the cosmological screening of the scalar field in high cosmological densities implies that the Hubble rate squared is always positive.
Another issue comes from the fact that, when the Hamiltonian becomes negative, the kinetic energy of K-mouflage excitations seen as particles may destabilize the vacuum and lead to a large background of gamma rays. For canonical ghosts,  it is known that one cannot extend the validity of the models for energies larger than a few MeV \cite{Cline:2003gs}. Here we revisit this issue for K-mouflage models and find that the validity range of these theories is even more restricted to energies always less than 1 keV when ghosts are present. Models where the kinetic term keeps the
standard positive sign do not have ghosts and do not suffer from this problem.

A first type of screening which differs from K-mouflage is   the chameleon \cite{Khoury:2003aq,Khoury:2003rn}(and also the Damour-Polyakov \cite{Damour:1994zq}) mechanism. In these models, screened regions are such that the Newtonian potential of dense objects is larger than the value of the scalar field outside the object. A second type of screening is the Vainshtein mechanism \cite{Vainshtein:1972sx} which only occurs for noncanonical scalar fields, and where screened regions correspond to a spatial curvature larger than a critical value. The K-mouflage mechanism is also present for noncanonical theories and is effective in regions where the gravitational acceleration is larger than a critical value, in a way reminiscent of the MOND hypothesis \cite{Babichev:2011kq}. For this reason, K-mouflage models differ from General Relativity on large scales and are well suited to be tested cosmologically. The absence of convergence towards GR in the large distance regime differs drastically from what happens for chameleons. The same lack of convergence  is also there for models with the Vainshtein property. On the other hand, the K-mouflage and Vainshtein mechanisms differ locally in the vicinity of a dense object of mass $m$ where the distance below which General Relativity is recovered scales like $m^{1/2}$ and $m^{1/3}$, respectively. They also differ drastically cosmologically as the latter (Vainshtein) allows for screening of cosmological overdensities like galaxy clusters while the former (K-mouflage) does not screen such large-scale structures. Consequently the theories with the K-mouflage property essentially behave like linear theory with a time dependent Newton constant up to quasilinear scales. In this paper, we will focus on the background evolution and leave the properties of structure formation on large scales for a companion paper \cite{Brax:2014ab}.

We will concentrate on the cosmology of K-mouflage models and leave their gravitational properties for further work. At the background level and for K-mouflage models leading to the late time acceleration of the expansion of the Universe, we find that the Hubble rate can differ significantly from the one of $\Lambda$-CDM for moderate redshifts. This fact corresponds to the cosmological screening at high cosmological densities of the scalar field
whose dynamics play a role only when the density of matter is sufficiently small. In the very recent past, the models converge to a $\Lambda$-CDM behavior. This implies that the deviations from $\Lambda$-CDM are maximal for intermediate redshifts of the order $1\lesssim z \lesssim 5$. For these redshifts, the effective energy density of the scalar can become positive after being negative in the distant past. This implies in particular that the equation of state is not bounded from below, can be less than $-1$ and can even diverge. This is an artefact of the the definition of the equation of state with no consequence on the dynamics of the models as the total matter density is always positive. It is remarkable that this result, i.e. the divergence of the equation of state,  stands for all models even when the Big Bang Nucleosynthesis constraints \cite{Uzan:2010pm} on the variation of masses are applied. Moreover, when excluding the possibility of ghosts, the K-mouflage models always cross the phantom divide. A feature which is strikingly different from chameleonlike models \cite{Brax:2004qh} and Galileons \cite{Li:2013tda,Barreira:2013eea}.

In section II, we recall the  physical classification of screening models for theories with second order equations of motions. K-mouflage models are particular as screening only occurs when the gravitational acceleration is large enough.  In part III, we introduce the K-mouflage models. In section IV, we study the tracking properties of the cosmological background and apply the BBN bounds to K-mouflage. In part V, we study the expansion history of the models and we confirm it numerically in section VI. Finally we study the constraints provided by the gamma ray flux due to ghosts in section VII. We then conclude in section VIII.

\section{Screening Mechanisms}

\subsection{Background effects}

Screening mechanisms can be nicely classified (for a single nearly massless scalar field on very large scales) by considering theories with second order equations of motion only \cite{Khoury:2013tda}. In a given environment characterized by overdensities  in a sparse background, the scalar field takes a background value, $\varphi_0 (t)$, which can be different inside and outside an overdensity. Expanding to second order, the Lagrangian of the  fluctuations compared to the background $\delta \varphi= \varphi-\varphi_0(t)$ becomes
\be
{\cal L} = - \frac{Z(\varphi_0)}{2}(\partial\delta \varphi)^2 - \frac{m^2(\varphi_0)}{2} (\delta\varphi)^2 -\beta (\varphi_0) \frac{\delta \varphi}{M_{\rm Pl}} \delta \rho_m
\ee
where $\delta\rho_m$ is the change of the matter density compared to the background. Test  particles follow the geodesics of the total potential\footnote{This potential can be obtained by going to the Jordan frame where test particles interact with the Jordan metric. In the Jordan frame, there is a second Newtonian potential  $\Phi=\Psi_{\rm N}- {\beta(\varphi_0)} \frac{\delta \varphi}{M_{\rm Pl}}$. Notice that $\Phi+\Psi=2\Psi_{\rm N}$ implying that
the scalar field has no direct effect on lensing and only acts on the geodesics through $\Psi$.}
\be
\Psi=\Psi_{\rm N}+ {\beta(\varphi_0)} \frac{\delta \varphi}{M_{\rm Pl}} ,
\ee
where the Newtonian potential $\Psi_{\rm N}$ satisfies the Poisson equation
\be
\nabla^2 \Psi_{\rm  N} = 4\pi A(\varphi_0) \cG_{\rm N} \delta \rho_{\rm m}
\label{Poisson-1}
\ee
and $\delta\varphi$ is due to the presence of the overdensity. Notice that the true Newtonian constant, as defined from the Poisson equation, is not $\cG_{\rm N}= 1/(8\pi M_{\rm Pl}^2)$ but $A(\varphi_0) \cG_{\rm N}$ due to the time dependence of the background and the coupling
function $A(\varphi)$ characterizing the model. Similarly, we have
\be
\beta (\varphi) = M_{\rm Pl} \frac{\dd\ln A(\varphi)}{\dd\varphi}
\ee
for all of these models. The time dependence of Newton's constant has important consequences as it can lead to large modifications of Big Bang Nucleosynthesis (BBN) as well as changes in the orbits of planets and stars which can be constrained using binary pulsars and the Lunar Ranging experiment testing the strong equivalence principle, with a current bound of $\vert \frac{\dot \cG_{\rm N}}{\cG_{\rm N}}\vert =\vert \frac{\dot A}{A}\vert \le 0.2 H_0$ \cite{Babichev:2011iz}. The BBN constraint imposes that the overall variation of particle masses is less than ten percent since BBN \cite{Uzan:2010pm}, i.e $\vert \frac{\Delta A}{A}\vert \le 0.1$ where $\Delta A$ is the variation of A.
Thus, $A \simeq 1$ in most cosmological configurations in these models.

Screening is defined as a reduction of the effect of the scalar field from the free and linear case of a point particle coupled with a strength $\beta (\varphi_0)$ to matter, i.e. screening occurs when $\Psi$ is smaller than $ (1+2\beta^2 (\varphi_0)) \Psi_{\rm N}$. Of course this is not equivalent, especially when $\beta(\varphi_0)$ is large, to requiring that the effect of ${\beta(\varphi_0)} \frac{\delta \varphi}{M_{\rm Pl}}$ is small compared to $\Psi_{\rm N}$, which also must be investigated for each experimental situation depending on the appropriate bound on the correction to $\Psi_{\rm N}$.

\subsection{Chameleon and Damour-Polyakov}

Let us focus on theories with $Z(\varphi)=1$ first, i.e. canonically normalized scalars.
At the linear level and in the quasistatic approximation \cite{Brax:2004qh}, the equations of motion give  that
\be
\frac{\delta \varphi}{M_{\rm Pl}}= - \frac{\beta (\varphi_0)\delta \rho_m}{M_{\rm Pl}^2 \left(m^2(\varphi_0)+ \frac{k^2}{a^2} \right)} ,
\label{dphi-lin-1}
\ee
where $k=1/x$ is the comoving wave number of interest, and
\be
\Psi= \left[ 1 + \frac{2\beta^2(\varphi_0)}{ 1+ m^2(\varphi_0)a^2/k^2} \right] \, \Psi_{\rm N} .
\ee
This allows for a direct effect of modified gravity on linear cosmological scales as
General Relativity is recovered on very large scales outside the Compton radius,
$k/a \lesssim m(\varphi_0)$, and changes to gravity occur inside the Compton radius,
$k/a \gtrsim m(\varphi_0)$, with a strength of $(1+2 \beta^2(\varphi_0))$.
When overdensities become greater, the linear approximation is not valid anymore and screening occurs. From Eq.(\ref{dphi-lin-1}) and Poisson's equation (\ref{Poisson-1}), in the small-scale
linear regime we have $\delta\varphi/M_{\rm Pl} \simeq 2 \beta \Psi_{\rm N}$. Therefore,
the condition for the onset of nonlinear screening, $|\delta\varphi| \sim |\varphi_0|$, can also
be written as a condition on the value $\Psi_{\rm N}$ of the Newtonian potential of the object,
\be
\left| 2 \beta(\varphi_0) \Psi_{\rm N} \right| \gtrsim \left| \frac{\varphi_0}{M_{\rm Pl}} \right| .
\label{screen-1}
\ee
Local tests such as the screening of the Milky Way and Solar System experiments impose that the cosmological range of
the scalar interaction must be less than 1 Mpc, which implies that most of the cosmological effects of these models are on quasilinear scales.
On the other hand, to satisfy the screening condition (\ref{screen-1}) for objects such as the Earth with $|\Psi_\oplus|\sim 10^{-9}$, the field variation in the Milky Way is bounded by
$|2\beta M_{\rm Pl}\Psi_\oplus|$, which in turns gives a constraint on the variation of
$A(\varphi)$ as
\be
\left| \frac{\Delta A}{A} \right| \lesssim  \left| \beta (\varphi_0) \frac{\varphi_0}{M_{\rm Pl}} \right|
 \le \left| 2 \beta^2(\varphi_0) \Psi_{\oplus} \right| .
\ee
This implies that Newton's constant hardly changed since BBN.

\subsection{K-mouflage and Vainshtein mechanisms}

The situation changes for theories with nontrivial kinetic terms. At the linear level and again
in the quasistatic approximation, for models dominated by the kinetic term where we
can neglect the potential term $m^2 (\delta\varphi)^2/2$, we have that
\be
\frac{\delta \varphi}{M_{\rm Pl}}= - \frac{\beta (\varphi_0) a^2 \delta \rho_m}{M_{\rm Pl}^2 Z(\varphi_0) k^2} = \frac{2\beta}{Z} \, \Psi_{\rm N} ,
\label{dphi-Phin}
\ee
and therefore
\be
\Psi= \left[ 1 + \frac{2\beta^2(\varphi_0)}{Z(\varphi_0)} \right] \, \Psi_{\rm N} ,
\ee
and screening occurs when
\be
Z(\varphi_0)\gtrsim 1 .
\ee
Nonlinearly, the modification of gravity around an overdensity is still suppressed when $Z$ is large. We can expand to leading order
\be
Z(\varphi)= 1+ a(\varphi) \frac{(\partial \varphi)^2}{\cM^4} + b(\varphi) L^2 \frac{\Box \varphi}{M_{\rm Pl}}+\dots ,
\ee
where $\cM$ is a suppression scale characterizing the model, $L$ a typical length scale, and $a(\varphi)$ and $b(\varphi)$ two  functions of the order one. Cubic and higher order derivatives are forbidden as they would induce equations of motions
of the order larger than two. This implies that this leading order expansion of $Z$ captures the essence of the possible screening mechanisms for theories with second order equations of motion.

\subsubsection{Vainshtein scenario}

When $a=0$, the suppression of the scalar field effect is due to the Vainshtein effect when the highest derivative in $Z$ satisfies
\be
\frac{\vert \nabla^2  \varphi\vert}{M_{\rm Pl}} \gtrsim L^{-2} ,
\ee
implying from Eq.(\ref{dphi-Phin}) that a necessary condition for screening is
\be
\vert \nabla^2 \Psi_{\rm N} \vert \gtrsim \frac{\vert \nabla^2 \varphi\vert}{2\beta M_{\rm Pl}} \gtrsim \frac{1}{2\beta L^2}
\label{curv}
\ee
when $\beta$ is a slowly varying function of $\varphi$.
Therefore, instead of a criterion on the amplitude of the Newtonian potential itself,
as in Eq.(\ref{screen-1}) for
chameleon models, we now have a criterion on its curvature, which also reads as a condition
on the density through Poisson's equation, that must be greater than a critical value determined
by the length scale $L$.

Taking the Newtonian potential around a dense object of mass $m$ and understanding the condition (\ref{curv}) in the sense of distribution averaging over a ball of radius $r$,  we find that screening occurs inside the Vainshtein radius \cite{deRham:2012az}
\be
R_V = \left( \frac{3\beta L^2  m}{4\pi M^2_{\rm Pl} }\right)^{1/3} .
\ee
For quasilinear cosmological structures, with a density constrast of $\delta$, the Poisson
equation reads in the current Universe as
\be
\nabla^2 \Psi_{\rm N} = \frac{3}{2} A(\varphi_0)  \Omega_{\rm m} H_0^2 \delta ,
\ee
where $\Omega_{\rm m0}$ is the matter density cosmological parameter today. This can lead to screening when
\be
 3A(\varphi_0)  \Omega_{\rm m0}  H_0^2 \delta \gtrsim \frac{1}{\beta(\varphi_0) L^2}.
\ee
The most interesting models are the ones where the scale $L$ is the size of the current Universe $H_0^{-1}$. In massive gravity models, this corresponds to a mass of the graviton of the order $H_0$ and hence effects of modified gravity up to the largest scales.
In this case, screening  occurs when
\be
\delta \gtrsim \frac{1}{ 3\Omega_{\rm m 0} A(\varphi_0) \beta(\varphi_0)} ,
\label{delta-k-Vainshtein}
\ee
which is realized for all overdensities with a density contrast larger than a number of the order one as soon as $A(\varphi_0)\sim 1$ from the BBN constraint and $\beta(\varphi_0)$ is not too small (which would render the screening effect
irrelevant).
Hence all quasilinear structures in the Universe are screened in the Vainshtein models and nonlinear effects must be taken into account even on mildly nonlinear scales.

\subsubsection{K-mouflage scenario}

When $b=0$, the suppression of the scalar field effect is due to the K-mouflage  effect when
\be
\vert \nabla  \varphi\vert \gtrsim \cM^2 ,
\ee
implying from Eq.(\ref{dphi-Phin}) that a necessary condition for screening is
\be
\vert \nabla \Psi_{\rm N} \vert \gtrsim  \frac{\vert \nabla \varphi\vert}{2 \beta M_{\rm Pl}} \gtrsim  \frac{\cM^2}{2 \beta M_{\rm pl}} \label{grad}
\ee
when $\beta$ is a slowly varying function of $\varphi$.
Therefore, instead of a criterion on the amplitude of the gravitational potential, as for chameleon
 models in Eq.(\ref{screen-1}), or on its curvature, as for Vainshtein-mechanism models
 in Eq.(\ref{curv}), we now have a criterion on the gradient of the gravitational potential, that is,
 the gravitational acceleration.

Taking the Newtonian potential around a dense object of mass $m$, we find that screening occurs inside the K-mouflage radius \cite{Babichev:2009ee,Brax:2012jr}
\be
R_K = \left( \frac{\beta m}{4\pi  M_{\rm Pl} \cM^2} \right)^{1/2} .
\ee
For quasilinear cosmological structures, we find that screening occurs when the wave number $k$ characterizing a given structure satisfies
\be
k \lesssim 3 \Omega_{\rm m 0} A(\varphi_0) \beta(\varphi_0) \frac{H_0^2 M_{\rm Pl}}{\cM^2} \delta
\ee
Taking $\cM^4 \sim 3 \Omega_{\Lambda 0} M_{\rm Pl}^2 H_0^2$ to recover the acceleration
of the Universe now,  we have that
\be
\frac{k}{H_0}\lesssim \sqrt{\frac{3}{\Omega_{\Lambda 0}}} \Omega_{\rm m 0} A(\varphi_0) \beta(\varphi_0) \delta ,
\label{k-delta-screen}
\ee
which corresponds to superhorizon scales if $\delta \sim 1$.
As a result, quasilinear objects in the Universe are not screened in K-mouflage models.

For a given scale (such as $x \sim 0.1 - 1 h^{-1}$Mpc for large-scale structures and $k=1/x$),
Eq.(\ref{k-delta-screen}) actually means that only very dense regions, $\delta \gtrsim k/H_0$
are screened. In particular, the condition becomes more severe for smaller scales,
in contrast with the scale-independent condition (\ref{delta-k-Vainshtein}) found for
the Vainshtein scenario.

In the following we shall concentrate on the background cosmology of K-mouflage. We will find that in the early Universe where densities are large, the effects of the scalar field are screened. The study
of the effects of K-mouflage on structure formation is left for a companion paper
\cite{Brax:2014ab}.

\section{K-mouflage}
\label{K-mouflage}

\subsection{Definition of the model}
\label{definition-model}

We consider scalar field models where the action in the
Einstein frame has the form
\beqa
S & = & \int \dd^4 x \; \sqrt{-g} \left[ \frac{\MPl^2}{2} R + {\cal L}_{\varphi}(\varphi)
\right] \nonumber \\
&& + \int \dd^4 x \; \sqrt{-\tg} \, {\cal L}_{\rm m}(\psi^{(i)}_{\rm m},\tg_{\mu\nu}) ,
\label{S-def}
\eeqa
where
 $g$ is the determinant of the metric tensor $g_{\mu\nu}$, and
$\psi^{(i)}_{\rm m}$ are various matter fields.
The additional scalar field $\varphi$ is explicitly coupled to matter through the
Jordan-frame metric $\tg_{\mu\nu}$, which is given by the conformal rescaling
\beq
\tg_{\mu\nu} = A^2(\varphi) \, g_{\mu\nu} ,
\label{g-Jordan-def}
\eeq
and $\tg$ is its determinant.
We have already considered various canonical scalar field models in previous
works \cite{BraxPV2012,BraxPV2013}, with
${\cal L}_{\varphi} = - (\pl\varphi)^2/2 - V(\varphi)$.
In this paper, we consider models with a nonstandard kinetic term
\beq
{\cal L}_{\varphi}(\varphi) = \cM^4 \, K\left(\frac{X}{\cM^4}\right) \;\;\; \mbox{with} \;\;\;
X = - \frac{1}{2} \, \pl^{\mu}\varphi \, \pl_{\mu}\varphi .
\label{K-def}
\eeq
[We use the signature $(-,+,+,+)$ for the metric.]
To focus on the behaviors associated with the nonstandard kinetic term $K$, we
do not add a potential $V(\varphi)$ or a mixed dependence $K(\varphi,X)$ on
the field value and the derivative terms.
Here, $\cM^4$ is an energy scale that will be of the order of the current energy density,
(i.e., set by the cosmological constant), to recover the late time accelerated expansion
of the Universe. We can choose $\cM^4>0$ without loss of generality.

It is convenient to introduce the dimensionless variable $\chi$ by,
\beq
\chi = \frac{X}{\cM^4}  = - \frac{1}{2\cM^4} \; \pl^{\mu}\varphi\pl_{\mu}\varphi .
\label{chi-def}
\eeq
Then, the canonical behavior [i.e., $K \sim \chi \propto -(\pl\varphi)^2/2$],
with a cosmological constant $\rho_{\Lambda} = \cM^4$ [see Eq.(\ref{rho-phi-def})
below], is recovered at late time in the weak-$\chi$ limit if we
have:
\beq
\chi \rightarrow 0 : \;\;\; K(\chi) \simeq -1 + \chi + ... ,
\label{K-chi=0}
\eeq
where the dots stand for higher order terms. The minus sign of the constant $-1$
at $\chi=0$ is set by the condition $\rho_{\Lambda} = - \cM^4 K(0)>0$, i.e. the corresponding vacuum energy for a uniform configuration should be positive as
we have $\cM^4>0$.

The Klein-Gordon equation that governs the dynamics of the scalar field
$\varphi$ is obtained from the variation of the action (\ref{S-def}) with respect to
$\varphi$. This gives
\beq
\frac{1}{\sqrt{-g}} \pl_{\mu} \left[ \sqrt{-g} \; \pl^{\mu} \varphi \; K' \right] -
\frac{\dd\ln A}{\dd\varphi} \; \rho_E = 0 ,
\label{KG-1}
\eeq
where $\rho_E=- g^{\mu\nu}T_{\mu\nu}$ is the Einstein-frame matter density,
and we note with a prime $K'=\dd K/\dd\chi$.

\subsection{Positive- and negative-$\chi$ tails}
\label{chi-p-m}

From Eq.(\ref{chi-def}), we can already see that uniform-field configurations
have $\chi \geq 0$ whereas (quasi)static configurations have $\chi\leq 0$.
Thus, the evolution of the cosmological background, where all fields are uniform,
only involves the $\chi\geq 0$ part of the kinetic function $K(\chi)$.
As we show in a companion paper,
when we consider the formation of cosmological structures (the cosmic web,
pancakes, filaments, and clusters of galaxies), we remain rather close to the
background value $\bar{\chi}>0$.
It is only for very dense regions, where $\delta\rho/\bar{\rho} > ct/r$, that the
variable $\chi$ can significantly depart from the background value
and that we can reach the static limit (e.g., an isolated static field configuration
in a Minkowski-like background) where $\chi<0$.
More precisely, we find in the companion paper that on subhorizon scales we can
distinguish the two regimes
\beq
\frac{ctk}{a} \gg 1, \;\; \frac{\delta\rho}{\bar\rho} \ll \frac{ctk}{a} : \;\;\;
\delta\varphi \;\; \mbox{is linear} ,
\label{psiA-1}
\eeq
and
\beq
\frac{ctk}{a} \gg 1, \;\; \frac{\delta\rho}{\bar\rho} \gtrsim \frac{ctk}{a} :  \;\;\;
\delta\varphi \;\; \mbox{is nonlinear} ,
\label{psiA-2}
\eeq
for which the fluctuations of the scalar field remain small, i.e.
\beq
\frac{c t k}{a} \gg 1 : \;\;\; \frac{\delta\varphi}{\bar\varphi} \sim
\frac{a^2}{c^2t^2k^2} \, \frac{\delta\rho}{\bar\rho}
\ll \frac{\delta\rho}{\bar\rho} ,
\label{dphi-drho-2}
\eeq
where $k\equiv 1/x$ is the typical comoving wave number of interest.
In Eqs.(\ref{psiA-1}) and (\ref{psiA-2}), ``linear" and ``nonlinear'' denotes whether we can
linearize the Klein-Gordon equation (\ref{KG-1}) around the background value $\bar{\varphi}$
to obtain the large-scale cosmological fluctuations $\delta\varphi\equiv \varphi-\bar{\varphi}$
of the scalar field.

For instance, for a typical cluster of galaxies we have $\delta\rho/\bar\rho \sim 200$
and $ct_0/r \sim c/(H_0r) \sim 3000$ (with $r\sim 1 h^{-1}$Mpc), while for the
Solar System, up to the Jupiter orbit, we have
$\delta\rho/\bar\rho \sim 10^{20}$ and $ct_0/r \sim 10^{14}$.
Thus, for large-scale cosmological structures (the cosmic web and clusters) we
are in the small-scale (i.e., below the horizon or Hubble radius) and
moderate-density regime (\ref{psiA-1}),
whereas in the Solar System we are in the small-scale
high-density regime (\ref{psiA-2}).
Thus, as compared with the mean background density on the size of the Universe,
which corresponds to the overall geometry and dynamics of the Universe,
structures correspond to both smaller scales and higher densities, and different
regimes are reached depending on the relative magnitudes of the relevant density
and scale as compared with $\bar\rho$ and $c/H_0$. Hence moderate-density regions on large scales are described in the linear regime around the background $\bar \chi >0$. Much denser objects such as the Sun are in the nonlinear regimes where
the branch $\chi \le 0$ must be considered.

In summary, for the cosmological dynamics, both at the level of the background and of the
formation of the large-scale structures, we are in the regime associated with
$\chi>0$, whereas the Solar System corresponds to the static regime $\chi<0$
which  is far from the cosmological background.
In this paper, we focus on the background cosmological aspects of these modified gravity models.
In the companion paper, we also show that the nonlinearities of the kinetic function $K(\chi)$
do not give rise to different scale-dependent qualitative behaviors.
Below the horizon, in the regime (\ref{psiA-1}), the modifications of gravity only
appear as a time dependent and scale-independent modifications of the equations
of motion (e.g., of Newton's constant), with regard to
large-scale structure formation. This also means that we do not recover General
Relativity on large scales and that the K-mouflage radius is smaller than these
cosmological structures.
On the other hand, the Solar System corresponds to a different regime where
General Relativity may be recovered through the K-mouflage mechanism,
depending on the behavior of the kinetic function $K(\chi)$ for large negative
$\chi$.
This may allow one to build models that are consistent with current Solar System tests
of General Relativity.

Because these two regimes are well separated and correspond to different parts of
the kinetic function $K(\chi)$, which could behave in unrelated manners if we go
beyond polynomials that imply identical power laws at
$\pm\infty$, it is convenient to separate the analysis of these two regimes.
Thus, we only consider the cosmological part in this paper and the companion
paper, and we leave the study of higher-density regions, giving rise to the K-mouflage
mechanism, to a future paper.

\subsection{Specific models}
\label{Specific-models}

For numerical computations we need to specify the kinetic function $K(\chi)$.
As noticed above, the evolution of the cosmological background, where all fields are uniform,
and of cosmological large-scale structures in the regime (\ref{psiA-1}),
only involve the $\chi\geq 0$ part of the kinetic function $K(\chi)$.
Then, as can already be seen from Eq.(\ref{KG-1}), two different behaviors
can be obtained, depending on whether the derivative $K'(\chi)$ has a zero or
not on the positive semiaxis.
Because of the low-$\chi$ behavior (\ref{K-chi=0}), which gives $K'(0)=1$, the case
without zero-crossing implies the positive sign, $K'>0$ for $\chi \geq 0$.
As a result, we consider the following three simple examples,
\beqa
``\mbox{no-}\chi_* , \; K' \geq 1 " & : & \;\;   K(\chi) = -1 + \chi + K_0 \, \chi^m , \nonumber \\
&& K_0 > 0 ,  \;\; m \geq 2 ,
\label{K-power-1}
\eeqa
\beqa
``\mbox{with-}\chi_* , \; K' \leq 0 " & : & \;\;   K(\chi) = -1 + \chi + K_0 \, \chi^m , \nonumber \\
&& K_0 < 0 ,  \;\; m \geq 2 ,
\label{K-power-2}
\eeqa
and
\beq
``\mbox{with-}\chi_* , \; K' \geq 0 "  :  \;\;   K(\chi) = -1 + \chi - \chi^2 + \chi^3/4 .
\label{K-power-3}
\eeq
The first model (\ref{K-power-1}) corresponds to scenarios where $K'$ never comes across
a zero (``no-$\chi_*$'') during the background cosmological evolution and remains positive.
The second and third models correspond to scenarios where $K'$ comes across
a zero (``with-$\chi_*$'') at late times (in fact, at infinite time), from below [Eq.(\ref{K-power-2})]
or from above [Eq.(\ref{K-power-3})], as $\chi$ rolls down from $+\infty$.
Because the background value of the variable $\chi$ always remains above $\chi_*$,
in the last two cases it never reaches the low-$\chi$ regime (\ref{K-chi=0}) and we could have
omitted the term $-1$ [the finite cosmological constant arising from $K(\chi_*)$ instead of
$K(0)$].

More generally, the first two terms in Eq.(\ref{K-power-1}), $(-1+\chi)$, represent
the first order expansion over $\chi$ of a generic function $K(\chi)$, as in
Eq.(\ref{K-chi=0}), so that we recover a canonical scalar field with a cosmological
constant for small time and spatial gradients.
The third term $K_0 \, \chi^m$ represents the large-$\chi$ behavior of the function
$K(\chi)$, or more precisely the relevant exponent at the time of interest.
For numerical computations, we consider the low order cases $m=2$ and $3$ in
models (\ref{K-power-1}) and (\ref{K-power-2}).

For the coupling function $A(\varphi)$, we consider the simple power laws,
\beq
n \in {\mathbb N} , \;\; n \geq 1 : \;\;\;
A(\varphi) = \left( 1 + \frac{\beta\varphi}{n \MPl} \right)^n ,
\label{A-power-1}
\eeq
which include the linear case $n=1$,
and the exponential limit for $n \rightarrow +\infty$,
\beq
A(\varphi) = e^{\beta \varphi/\MPl} .
\label{A-exp-1}
\eeq
Without loss of generality, we normalized the field $\varphi$ (by the appropriate
additive constant) so that $A(0)=1$.

These forms ensure that $A(\varphi)$ and $K(\chi)$ are always well defined, for
all values of the fields.
The action (\ref{S-def}) is invariant with respect to the transformation
$(\varphi,\beta) \rightarrow (-\varphi,-\beta)$; therefore we can choose $\beta >0$.
Thus, in addition to the usual cosmological parameters, our system is defined by the
five parameters
\beq
\{\beta, n ; K_0 , m ; \cM^4 \} \;\; \mbox{with} \;\; \beta > 0, \; \cM^4 > 0 , \;
n \geq 1 , \; m \geq 2 ,
\label{param-def}
\eeq
except for the model (\ref{K-power-3}) where there are no parameters $\{K_0,m\}$ as the
kinetic function is fixed.
The scale $\cM$ is not an independent parameter. For a given value of the set
$\{ \beta, n ; K_0 , m \}$ and of $H_0$, it is fixed by the value of $\Omega_{\rm m0}$
today.
Thus, in the numerical computations below, we choose the same set of cosmological
parameters today, given by the Planck results \cite{Planck-Collaboration:2013}.
Then, for any set $\{ \beta, n ; K_0 , m \}$, we tune $\cM$ to obtain the observed
dark energy density today. As noticed above, this means that
$\cM^4 \sim \bar\rho_{\rm de 0}$.

\section{Cosmological background}
\label{background}

\subsection{Equations of motion}
\label{Equations-of-motion}

We focus on the matter era and we only consider nonrelativistic pressureless matter
and the scalar field $\varphi$.
Because the gravitational Lagrangian is not modified in the action (\ref{S-def}),
the metric $g_{\mu\nu}$ and the Einstein tensor associated with the homogeneous
and isotropic background follow the usual FLRW form,
\beq
g_{\mu\nu}  = \mbox{diag}(-1,a^2,a^2,a^2) ,
\label{gmunu-back-def}
\eeq
\beq
G_{00} = 3H^2, \;\;\; G_{ij} = -a^2(2\dot{H}+3H^2)  \delta_{i,j} ,
\label{G00-back-def}
\eeq
where $a(t)$ is the scale factor, $H(t)=\dot{a}/a$ the Hubble expansion rate,
$\delta_{i,j}$ the Kronecker symbol, and $i,j=1,2,3$.
Then, the Einstein equations lead to the usual Friedmann equations,
\beqa
3 \MPl^2 H^2 & = & \bar{\rho}_E + \bar{\rho}_{\varphi} ,
\label{Friedmann-1} \\
-2 \MPl^2 \dot{H} & = & \bar{\rho}_E + \bar{\rho}_{\varphi} + \bar{p}_{\varphi}
\label{Friedmann-2}
\eeqa
where $\rho_E$, $\rho_{\varphi}$ and $p_{\varphi}$, are the matter and scalar
field energy densities and pressure (in the Einstein frame).
For the Lagrangian (\ref{K-def}), the latter read as
\beq
\bar{\rho}_{\varphi} = - \cM^4 \bar{K} + \dot{\bar\varphi}^2 \, \bar{K}'  , \;\;\;
\bar{p}_{\varphi} = \cM^4 \bar{K} .
\label{rho-phi-def}
\eeq
Here and in the following, the overbar denotes uniform background quantities,
and the dimensionless field $\chi$ defined in Eq.(\ref{chi-def}) simplifies as
\beq
\bar{\chi} = \frac{\dot{\bar\varphi}^2}{2\cM^4} .
\label{chi-bar-def}
\eeq

Because the matter Lagrangian involves the rescaled metric
$\tilde{g}_{\mu\nu}$, its energy-momentum tensor $T_{(\rm m)}$ is no longer conserved,
$D_{\mu} T^{\mu\nu}_{(\rm m)} \neq 0$, where $D_{\mu}$ is the Einstein-frame covariant
derivative.
Nevertheless, the Jordan-frame
energy-momentum tensor still satisfies the standard conservation law
$\tilde{D}_{\mu} \tilde{T}^{\mu\nu}_{(\rm m)} = 0$. Using Eq.(\ref{g-Jordan-def}),
this leads to
\beq
\dot{\bar{\rho}}_E = - 3 H \, \bar{\rho}_E + \frac{\dd\ln \bar{A}}{\dd\bar\varphi} \,
\dot{\bar\varphi} \, \bar{\rho}_E ,
\label{rhoE-conserv}
\eeq
where we have used the relation $\rho_J=A^{-4}\rho_E$ between the Jordan-frame and
Einstein-frame matter densities. It is convenient to introduce the rescaled matter
density $\rho$,
\beq
\rho = A^{-1} \rho_E , \;\;\; \mbox{hence} \;\;\;
\dot{\bar{\rho}} = - 3 H \, \bar{\rho} \;\; \mbox{and} \;\;
\bar\rho = \frac{\bar\rho_0}{a^3} ,
\label{rho-conserv}
\eeq
which satisfies the standard conservation equation
(and $\bar\rho_0$ is the background density $\bar\rho$ today).

On the other hand, the Klein-Gordon equation (\ref{KG-1}) reads as
\beq
\pl_t \left( a^3 \dot{\bar\varphi} \bar{K}' \right) =
- \frac{\dd \bar{A}}{\dd\bar\varphi} \, \bar{\rho} \, a^3 ,
\label{KG-2}
\eeq
or equivalently
\beq
\ddot{\bar\varphi} \left( \bar{K}' + \frac{\dot{\bar\varphi}^2}{\cM^4} \,
\bar{K}'' \right)
+ 3 H \, \dot{\bar\varphi} \, \bar{K}' = - \frac{\dd \bar{A}}{\dd\bar\varphi} \,
\bar{\rho} .
\label{KG-3}
\eeq
In particular, this leads to the modified conservation equation
\beq
\dot{\bar\rho}_{\varphi} = - 3 H ( \bar{\rho}_{\varphi} + \bar{p}_{\varphi} )
- \frac{\dd \bar{A}}{\dd\bar\varphi} \, \bar{\rho} \, \dot{\bar\varphi} ,
\label{rho-phi-no-conserv}
\eeq
where the exchange term between the matter and scalar field compensates
as it should the last term of Eq.(\ref{rhoE-conserv}) [because the full
energy-momentum tensor must be conserved in the Einstein frame,
$D_{\mu} ( T^{\mu\nu}_{(\rm m)} + T_{(\varphi)}^{\mu\nu} ) = 0$].
As for the matter sector, it is convenient to introduce an effective scalar field energy
density,
\beq
\rho_{\varphi}^{\rm eff} = \rho_{\varphi} + [ A(\varphi)-1] \rho ,
\label{rho-phi-eff-def}
\eeq
which satisfies the standard conservation equation (the pressure $p_{\varphi}$
is not modified)
\beq
\dot{\bar{\rho}}_{\varphi}^{\rm eff} = - 3 H (\bar{\rho}_{\varphi}^{\rm eff}
+ \bar{p}_{\varphi} ) .
\label{conserv-1}
\eeq
Then, the Friedmann equations (\ref{Friedmann-1})-(\ref{Friedmann-2}) also
read as
\beqa
3 \MPl^2 H^2 & = & \bar{\rho} + \bar{\rho}_{\varphi}^{\rm eff} ,
\label{Friedmann-3} \\
-2 \MPl^2 \dot{H} & = & \bar{\rho} + \bar{\rho}_{\varphi}^{\rm eff} + \bar{p}_{\varphi} ,
\label{Friedmann-4}
\eeqa
and we define the time dependent cosmological parameters
\beq
\Omega_{\rm m} = \frac{\bar\rho}{\bar\rho+\bar{\rho}_{\varphi}^{\rm eff}} , \;\;
\Omega_{\varphi}^{\rm eff} = \frac{\bar{\rho}_{\varphi}^{\rm eff}}
{\bar\rho+\bar{\rho}_{\varphi}^{\rm eff}} , \;\;
w_{\varphi}^{\rm eff} = \frac{\bar{p}_{\varphi}}{\bar{\rho}_{\varphi}^{\rm eff}} .
\label{w-def}
\eeq

\subsection{Recovery of the matter era at early times}
\label{early-matter-era}

To obtain a realistic cosmology, we must check that we recover the usual
matter-dominated expansion at early time (we only consider the matter era in this
paper), that is, we must have $\bar\rho_{\varphi}^{\rm eff} \ll \bar\rho$ and
$\bar{p}_{\varphi} \ll \bar\rho$ as $t\rightarrow 0$.
From Eq.(\ref{rho-phi-eff-def}) this implies that $(\bar{A}-1) \rightarrow 0$,
hence $\bar\varphi \rightarrow 0$,
\beq
t \rightarrow 0 :  \;\; \bar\varphi \rightarrow 0 , \;\;
\bar{A} \simeq 1 + \frac{\beta\bar{\varphi}}{\MPl} , \;\;
\frac{\dd \bar{A}}{\dd\bar{\varphi}} \simeq \frac{\beta}{\MPl}  .
\label{phi-A-t0}
\eeq
Then, the Klein-Gordon equation (\ref{KG-2}) gives
$\dot{\bar\varphi}\bar{K}' \rightarrow \infty$.
We consider models such as (\ref{K-power-1})-(\ref{K-power-3})
with a power-law behavior at large positive $\chi$, hence
\beq
t \rightarrow 0 :  \;\; \bar{\chi} \rightarrow +\infty ,
\;\;\bar{K} \simeq K_0 \, \bar{\chi}^m .
\eeq
[The model (\ref{K-power-3}) has $K_0=1/4$ and $m=3$.]
Then, the Klein-Gordon equation (\ref{KG-2}) can be integrated as
\beq
t \rightarrow 0 : \;\; a^3 \dot{\bar\varphi} K_0 m
\left( \frac{\dot{\bar\varphi}^2}{2\cM^4} \right)^{m-1}
= - \frac{\beta \bar{\rho}_0}{\MPl} (t + \gamma) ,
\label{phib-1}
\eeq
where we used the conservation equation (\ref{rho-conserv}) and $\gamma$ is
an integration constant.
This gives $\dot{\bar\varphi} \sim [(t+\gamma)/t^2]^{1/(2m-1)}$, using
$a(t) \sim t^{2/3}$, and $\bar\rho_{\varphi} \sim [(t+\gamma)/t^2]^{2m/(2m-1)}$.
Requiring that $\bar\rho_{\varphi} \ll \bar\rho \sim t^{-2}$ for $t\rightarrow 0$
leads to:
\beq
\gamma=0 \;\;\; \mbox{and} \;\;\; m>1 .
\label{gamma-0}
\eeq
Therefore, the integration constant $\gamma$ vanishes and the Klein-Gordon
equation (\ref{KG-2}) can be integrated once to read as
\beq
t \geq 0 : \;\; a^3 \dot{\bar\varphi} \bar{K}' = - \int_0^t \dd t' \; \bar\rho_0
\frac{\dd\bar{A}}{\dd\bar\varphi}(t') .
\label{KG-int-1}
\eeq
The boundary condition (\ref{phi-A-t0}), $\bar\varphi \rightarrow 0$, also
implies
\beq
t \geq 0 : \;\; \bar\varphi = \int_0^t \dd t' \, \dot{\bar\varphi}(t') .
\label{phi-phip-1}
\eeq
This gives the early time power-law behaviors
\beqa
t \rightarrow 0 & : & |\bar\varphi| \sim  t^{2(m-1)/(2m-1)} , \;\;
|\dot{\bar\varphi}| \sim t^{-1/(2m-1)} , \nonumber \\
&& \bar\rho_{\varphi}^{\rm eff} \sim \bar\rho_{\varphi} \sim \bar{p}_{\varphi}
\sim t^{-2m/(2m-1)} ,
\label{phi-t-0}
\eeqa
and all kinetic models (\ref{K-power-1})-(\ref{K-power-3}) satisfy
$\bar\rho_{\varphi}^{\rm eff} \ll \bar\rho$ at early time, because $m>1$.
Moreover, we have
\beqa
t \rightarrow 0 & : & \bar\rho_{\varphi} \sim (2m-1) \cM^4 \bar{K}, \;\;
\bar\rho_{\varphi}^{\rm eff} \sim - \frac{2m-1}{m-1} \cM^4 \bar{K}, \nonumber \\
&& w_{\varphi}^{\rm eff} \simeq - \frac{m-1}{2m-1} .
\label{w-phi-eff-1}
\eeqa
The signs of $\bar\rho_{\varphi}$ and $\bar\rho_{\varphi}^{\rm eff}$ depend on the
sign of $K_0$. From Eqs.(\ref{KG-int-1})-(\ref{phi-phip-1}), the signs of
$\bar\varphi$ and $\dot{\bar\varphi}$ are opposite to the sign of $K_0$
(because $\beta>0$),
\beq
t \rightarrow 0 : \;\; K_0 \dot{\bar\varphi} < 0 , \;\; K_0 \bar\varphi < 0 .
\label{K0-sign-phi}
\eeq
This implies that in the high-density regime of the early Universe, the effects of the scalar field are screened.

\subsection{Classical stability of the background}
\label{classical-stability}

We discuss here the stability of the early time background solution obtained above when a small perturbation is initially applied.
The power-law solution obtained from Eq.(\ref{phib-1}), with the integration constant
$\gamma=0$, is the adequate solution of the original Klein-Gordon equation
(\ref{KG-2}), which reads in this regime as
\beq
\frac{\dd}{\dd t} \left( a^3 \dot{\bar\varphi}^{2m-1} \right) = - \frac{\beta}{M_{\rm Pl}} \bar{\rho}_0
\frac{(2\cM^4)^{m-1}}{K_0 m} .
\label{KG-m}
\eeq
If we write the background as $\bar\varphi = \bar\varphi_0 + \delta\bar\varphi$,
where $\bar\varphi_0$ is the peculiar solution obtained in
Sec.~\ref{early-matter-era}, we obtain at linear order
\beq
\frac{\dd}{\dd t} \left( a^3 \dot{\bar\varphi}_0^{2m-2} \delta\dot{\bar\varphi} \right) = 0 .
\eeq
Using the early time power-law behaviors $a \propto t^{2/3}$ and
$\dot{\bar\varphi}_0 \propto t^{-1/(2m-1)}$, from Eq.(\ref{phi-t-0}),
this yields
\beq
\delta\dot{\bar\varphi} \sim t^{-2m/(2m-1)}, \;\;\; \delta\bar\varphi \sim \mbox{constant} ,
\eeq
hence
\beq
\frac{\delta\dot{\bar\varphi}}{\dot{\bar\varphi}_0} \sim t^{-1} , \;\;\;
\frac{\delta\bar\varphi}{\bar\varphi_0} \sim t^{-2(m-1)/(2m-1)} .
\label{stab1}
\eeq
Therefore, the background solution $\bar\varphi_0$ obtained in Sec.~\ref{early-matter-era}
is stable and is a tracker solution for K-mouflage.

\subsection{Comparison with other modified-gravity scenarios}
\label{Comparison}

We can now compare the type of background solution in K-mouflage models and in chameleonlike and Galileon (as an example of the Vainshtein mechanism) models.
We have just seen that deviations from $\Lambda$-CDM are minimal in the early Universe and at late time for K-mouflage models. The latter requirement is phenomenological and must be imposed to retrieve
an equation of state close to -1 since $z\sim 1$. As a result, modifications to the Hubble rate can and do occur for intermediate values of $1\lesssim z \lesssim 5$. For chameleonlike models, the field tracks the minimum of the effective potential when the mass of the scalar is sufficiently larger than the Hubble rate \cite{Brax:2004qh}. In this case the deviations from $\Lambda$-CDM are minimal and of order $H^2/m^2$ which must be always less than $10^{-6}$ to comply with the absence of deviations from GR in the solar system. Galileons also have a tracker behavior \cite{Li:2013tda}, where $\dot \varphi \sim H_0^2/H$ and $H^2/H_0^2= \frac{1}{2} (\Omega_{\rm m0} a^{-3} + \sqrt{ \Omega_{\rm m0}^2 a^{-6} + 4\Omega_{\Lambda 0}})$ corresponding to an effective time dependent cosmological constant which grows with time.
In contrast, $\dot{\bar\varphi}$ and the dark energy density decrease with time in K-mouflage models. This also implies that the departures from the $\Lambda$-CDM
reference decrease more slowly at higher redshift for the K-mouflage models.

 In a similar fashion, in $f(R)$ theories of the form \cite{Hu:2007nk}
 $f(R) = -2\Lambda- f_{R_0}c^2 R_0^{n+1}/(nR^n)$, with $n>0$ (in the large-curvature regime),
 the cosmology becomes increasingly close to a $\Lambda$-CDM scenario at high redshift.
 For the K-mouflage scenario, the dark energy component does not converge to a cosmological
 constant at early times. Thus, $w^{\rm eff}_{\varphi}$ goes to a constant that is different from both
 $0$ and $-1$, see Eq.(\ref{w-phi-eff-1}), and the scalar field energy densities
 $\bar\rho_{\varphi}$ and $\bar\rho_{\varphi}^{\rm eft}$ actually keep growing, see
 Eq.(\ref {phi-t-0}). Then, we never recover a truly $\Lambda$-CDM behavior, with a constant
 dark energy density, but only the Einstein-de Sitter asymptotics because the dark energy
 density does not grow as fast as the matter density with redshift.

\subsection{BBN constraint}
\label{BBN}

Particle masses in the Einstein frame are given by
\beq
m_\psi= A(\varphi) m_\psi^0 ,
\label{m-psi}
\eeq
where $m_\psi^0$ is the bare mass appearing in the Lagrangian. As such, this implies
that masses become environmentally dependent.
Then, particle masses in the background density have changed since BBN by
an amount
\beq
\frac{\Delta m_{\psi}}{m_\psi^0} = \Delta \bar{A} .
\label{Delta-m-A}
\eeq
This corresponds to particles in moderate-density fluctuations, such as
Lyman-alpha clouds, filaments, and outer radii of X-ray clusters.
The BBN constraint imposes that particle masses must vary by less than
${\cal O} (30)\%$ \cite{Uzan:2010pm} and therefore $|\Delta\bar{A}| \lesssim 1$.
High-density regions, where the nonlinear K-mouflage mechanism ensures
a convergence to General Relativity through a vanishing of the scalar field gradients,
can display a local value $\varphi$ that differs from the background $\bar\varphi$.
Therefore, the criterium $|\Delta\bar{A}| \lesssim 1$ may not be sufficient
for some models, but it remains a necessary condition.

For the models (\ref{A-power-1})-(\ref{A-exp-1}) that we consider in this paper,
the BBN constraint $|\Delta\bar{A}| \lesssim 1$ reads as
\beq
\left| \frac{\beta\bar\varphi}{M_{\rm Pl}} \right| \lesssim 1 ,
\label{BBN-phi-1}
\eeq
as $\bar\varphi \rightarrow 0$ for $t\rightarrow 0$.

\section{Expansion history}
\label{expansion-history}

From the integrated form (\ref{KG-int-1}) of the Klein-Gordon equation we can see that
at late times, in the dark energy era, we have
\beq
t \rightarrow \infty : \;\;\; \dot{\bar\varphi} \bar{K}' \rightarrow 0.
\label{late-time-behavior}
\eeq
Therefore, we obtain two different behaviors, depending on whether $K'$ has a zero $\chi_*$
on the positive axis, which will set the large-time dynamics, or not, in which case
$\dot{\bar\varphi}$ and $\bar{\chi}$ go to zero.

\subsection{Expansion history for models with $K'>0$ for $\chi\geq 0$}
\label{expansion-history-K0-p}

We first consider the kinetic models (\ref{K-power-1}) with $K_0>0$, or
more generally kinetic functions such that $K'>0$ for $\chi \geq 0$
(i.e., no zero on the positive semiaxis).
This implies that $\bar{K}'$ runs from $+\infty$ to $1$ as $\bar\chi$ rolls down from
$+\infty$ to $0$ and $\dot{\bar\varphi}$ goes from $-\infty$ to $0$, following
the Klein-Gordon equation (\ref{KG-int-1}).
We obtain at late times
\beqa
t \rightarrow \infty & :  &  \bar\rho_{\varphi}^{\rm eff} \simeq \cM^4 , \;\;
a(t) \sim e^{\cM^2 t/(\sqrt{3}M_{\rm Pl})} , \nonumber \\
&& \bar\rho \propto a^{-3} , \;\;
|\dot{\bar\varphi}| \sim t \; a^{-3} , \;\;
\bar\varphi \rightarrow \mbox{constant} < 0 , \;\;\;\;\;\;
\label{a-exp-L}
\eeqa
and we recover a cosmological constant behavior, with a constant dark energy
density $\bar\rho_{\rm de} = \cM^4$.

Because the effective energy density $\bar\rho_{\varphi}^{\rm eff}$ is negative
at early times from Eq.(\ref{w-phi-eff-1}), it must change sign and vanish at a time,
$t_{\rm eff}$, before reaching the cosmological constant regime.
This implies an effective equation of state parameter $w_{\varphi}^{\rm eff}$ that
diverges at $t_{\rm eff}$.
To satisfy observational constraints, this time $t_{\rm eff}$ must occur sufficiently
far in the past, so that at low redshifts $z \lesssim 1$ where
$\bar\rho_{\varphi}^{\rm eff}$ becomes of the order $\rhob$ we are close to the
cosmological constant regime with $w_{\varphi}^{\rm eff} \simeq -1$.

Depending on the value of the parameters $K_0$ and $\beta$, the Universe can
go through different regimes.
To simplify the analysis, we can use the approximations
\beq
t_{\rm de} \sim t_0 , \;\; \cM^4 \sim \bar\rho_0 , \;\; a(t) \sim (t/t_0)^{2/3} , \;\;
\MPl^2 \sim \bar\rho_0 t_0^2 ,
\label{t0-approx-1}
\eeq
because the accelerated expansion of the Universe only started at a late time
$t_{\rm de}$, with $z_{\rm de} \lesssim 1$. Here $t_0$ is the current age of the Universe.
From the BBN constraint (\ref{BBN-phi-1}), we can write
$\dd\bar{A}/\dd\bar\varphi \simeq \beta/\MPl$. Then, the Klein-Gordon
equation (\ref{KG-int-1}) and Eq.(\ref{phi-phip-1}) give
$\dot{\bar\varphi} \bar{K}' \sim - \beta \rhob t/M_{\rm Pl}$
and
\beq
0 \leq t \leq t_0  : \;\; \frac{\beta\bar\varphi}{M_{\rm Pl}} \sim
- \frac{\beta^2}{\bar{K}'} .
\label{beta-phi-1}
\eeq
Thus, the BBN constraint (\ref{BBN-phi-1}) also implies that
\beq
\left| \frac{\beta^2}{\bar{K}'} \right| \lesssim 1 .
\label{BBN-2}
\eeq

For large $\bar\chi$, where $\bar{K}' \simeq K_0 m \bar\chi^m$, the
Klein-Gordon equation (\ref{KG-int-1}) [i.e., Eq.(\ref{phib-1}) with $\gamma=0$]
gives
\beq
t < t_m : \;\; \bar{\chi} \simeq
\left( \frac{\beta \bar\rho t}{K_0 m \MPl \sqrt{2\cM^4}} \right)^{2/(2m-1)} ,
\label{chi-m-1}
\eeq
whereas at later time, when $\bar{K}' \simeq 1$, we have
\beq
t > t_m : \;\; \bar{\chi} \simeq
\left( \frac{\beta \bar\rho t}{\MPl \sqrt{2\cM^4}} \right)^{2} .
\label{chi-1-1}
\eeq
The transition time $t_m$ is set by $mK_0 \bar\chi^{m-1} = 1$, hence
\beq
t_m = K_0^{1/[2(m-1)]} \beta \, t_ 0  .
\label{tm-def}
\eeq
[Here and in the following relations, we only look for scalings and orders of
magnitude, and we discard irrelevant factors of order unity such as $\sqrt{2}$
or $m$.]
Before this transition time, the scalar field $\bar\varphi$ shows the power-law
growth (\ref{phi-t-0}), and more precisely we obtain:
\beq
t < t_m : \frac{\beta\bar\varphi}{\MPl} \sim -
\left( \frac{\beta^{2m} t^{2(m-1)}}{K_0 t_0^{2(m-1)}} \right)^{1/(2m-1)} .
\label{beta-phi-power}
\eeq

\subsubsection{$K_0 \ll 1$}
\label{K0-small}

Let us first consider the case of $K_0 \ll 1$. Then, the Universe goes through
four stages (we focus on late times after matter-radiation equality):
\beqa
\rm{(I)} \!: && \!\!\! t < t_m , \; \bar{K} \simeq K_0 \bar{\chi}^m , \;
\frac{\bar\rho_{\varphi}^{\rm eff}}{\bar\rho} \sim t^{2(m-1)/(2m-1)} , \;\;\;\;
\label{K0-small-I}
\\
\rm{(II)} \! : && \!\!\! t_m < t < t_{\Lambda} , \; \bar{K} \simeq \bar{\chi} , \;
\frac{\bar\rho_{\varphi}^{\rm eff}}{\bar\rho} \simeq \mbox{constant} ,
\\
\rm{(III)} \! : && \!\!\! t_{\Lambda} < t  < t_{\rm de} , \; \bar{K} \simeq -1 , \;
\bar\rho_{\varphi}^{\rm eff} \simeq \cM^4 \ll \bar\rho  ,
\\
\rm{(IV)} \! : && \!\!\! t_{\rm de} < t , \; \bar{K} \simeq -1 , \;
\bar\rho_{\varphi}^{\rm eff} \simeq \cM^4 \gg  \bar\rho  ,
\label{K0-small-IV}
\eeqa
where we have used the result (\ref{phi-t-0}). The first three epochs are distinguished by
the regime of the kinetic function $K(\chi)$. The third and fourth epochs, where the
scalar field acts as a cosmological constant, are distinguished by the relative
importance of the matter and dark energy densities.
In practice, as explained above we can take $t_{\rm de} \sim t_0$.
To obtain a $\Lambda$-CDM-like behavior, we must also have
$t_{\Lambda} \lesssim t_0$ to recover $w_{\varphi}^{\rm eff} \simeq -1$ at late times.
This time $t_{\Lambda}$ corresponds to $\bar\chi=1$. Using Eq.(\ref{chi-1-1}),
we obtain
\beq
K_0 \ll 1 : \;\;\; t_{\Lambda} = \beta \, t_0 \;\;\; \mbox{hence} \;\;\; \beta \lesssim 1.
\label{t-Lambda-K0-small}
\eeq
The constraint $t_{\rm de} \simeq t_0$ implies
$\cM^4 \simeq \bar\rho_0$.

From Eqs.(\ref{tm-def}) and (\ref{t-Lambda-K0-small}), we have in this case the
ordering $t_m \ll t_{\Lambda} \lesssim t_0$.
In the time range $t_m < t \leq t_0$, the Klein-Gordon equation (\ref{KG-int-1})
gives $\dot{\bar{\varphi}} \sim - \beta \bar\rho_0 t_0^2/(M_{\rm Pl} t) \propto 1/t$.
Therefore, neglecting logarithmic corrections, we obtain
$\bar{\varphi} \sim \bar\varphi(t_m)$ and
\beq
t_m < t \leq t_0 : \;\; \frac{\beta\bar\varphi}{M_{\rm Pl}} \sim - \beta^2 .
\label{beta-phi-constant}
\eeq
From the constraint (\ref{t-Lambda-K0-small}), $\beta \lesssim 1$,
we can see that the BBN constraint (\ref{BBN-phi-1}) is also satisfied.

For very small $K_0$, the nonstandard kinetic term $K_0 \chi^m$ only plays
a role at early times before the dark energy component takes over. Then, it has
no impact on the expansion history of the Universe.

\subsubsection{$K_0 \gg 1$}
\label{K0-large}

Let us now turn to the case of $K_0 \gg 1$. Then, the kinetic term directly
shifts from the regime (I) to (III), that is, from $\bar{K} \simeq K_0 \bar\chi^m$ to
$\bar{K} \simeq -1$, and we have the sequence
\beqa
\rm{(I)} \!: && \!\!\! t < t_{\Lambda} , \; \bar{K} \simeq K_0 \bar{\chi}^m , \;
\frac{\bar\rho_{\varphi}^{\rm eff}}{\bar\rho} \sim t^{2(m-1)/(2m-1)} , \;\;\;\;\;\; \label{I-p}
\\
\rm{(II)} \! : && \!\!\! t_{\Lambda} < t  < t_{\rm de} , \; \bar{K} \simeq -1 , \;
\bar\rho_{\varphi}^{\rm eff} \simeq \cM^4 \ll \bar\rho  , \label{II-p}
\\
\rm{(III)} \! : && \!\!\! t_{\rm de} < t , \; \bar{K} \simeq -1 , \;
\bar\rho_{\varphi}^{\rm eff} \simeq \cM^4 \gg  \bar\rho  . \label{III-p}
\eeqa
(In other words, we now have $t_m>t_{\Lambda}$.)
The time $t_{\Lambda}$ now corresponds to $K_0 \bar\chi^m=1$. This gives
\beq
K_0 \gg 1 : \;\;\; t_{\Lambda} = K_0^{-1/(2m)} \, \beta \, t_0 \;\;\; \mbox{hence} \;\;\;
\beta \lesssim K_0^{1/(2m)} .
\label{t-Lambda-K0-large}
\eeq

There are  two possible orderings for the time $t_m$ which need to be discussed.
In the first case, $t_0 < t_m$, $\bar{K}' \simeq m K_0 \bar\chi^{m-1}$ until the present
time. This corresponds to $K_0^{1/[2(m-1)]} \beta > 1$ from Eq.(\ref{tm-def}), and
we obtain today:
\beqa
K_0^{1/[2(m-1)]} \beta > 1 , \;\;  t= t_0 & : & \nonumber \\
&& \hspace{-2.5cm} \bar{K}' \sim \left( \beta^{2m-2} K_0 \right)^{1/(2m-1)} > 1 ,
\nonumber \\
&& \hspace{-2.5cm} \frac{\beta\bar\varphi}{M_{\rm Pl}} \sim
- \left( \frac{\beta^{2m}}{K_0} \right)^{1/(2m-1)} \lesssim 1 ,
\label{K0-p-betaphi}
\eeqa
where we have used the constraint (\ref{t-Lambda-K0-large}).
In the second case, $t_m < t_0$, $\bar{K}' \simeq 1$ today. This corresponds to
$K_0^{1/[2(m-1)]} \beta < 1$ and we obtain:
\beqa
K_0^{1/[2(m-1)]} \beta < 1 , \;\;  t= t_0 & : & \nonumber \\
&& \hspace{-2cm} \bar{K}' \simeq 1 ,  \;\;
\frac{\beta\bar\varphi}{M_{\rm Pl}} \sim - \beta^2 \ll 1 ,
\label{K0-p-betaphi-2}
\eeqa
where we have used $\beta^2 < K_0^{-1/(m-1)} \ll 1$.
Therefore, the BBN constraint (\ref{BBN-phi-1}) is satisfied as soon as the
condition (\ref{t-Lambda-K0-large}) is verified.
The second case (\ref{K0-p-betaphi-2}), where $\beta$ is very small,
$\beta \lesssim K_0^{-1/[2(m-1)]} \ll 1$, can only give very small deviations from
a uniform quintessence model as it yields $|\beta\bar\varphi/M_{\rm Pl}| \ll 1$.
Indeed, this implies a coupling function $A(\varphi) \simeq 1$ that is almost
constant so that the Einstein-frame and Jordan-frame metrics are almost
identical [in scenarios with $\beta\sim 1$, the main effect of the scalar field
on the matter geodesics does not arise from the contributions of the
fluctuations of $\rho_{\varphi}$ to the metric but from the conformal rescaling
(\ref{g-Jordan-def}), see the study of the formation of large-scale structures in the
companion paper].

\subsubsection{$K_0 \sim 1$}
\label{K0-unity}

The case $K_0 \sim 1$ is the transition between the two previous scenarios.
We now have:
\beq
K_0 \sim 1 : \;\;\; t_ m = t_{\Lambda} = \beta \, t_0 \;\;\; \mbox{hence} \;\;\;
\beta \lesssim 1,
\label{t-Lambda-K0-one}
\eeq
and the Universe goes through three stages as in Eqs.(\ref{I-p})-(\ref{III-p}).
We also have $\bar{K}' \simeq 1$ today and the condition (\ref{t-Lambda-K0-one})
also implies that the BBN constraint (\ref{BBN-phi-1}) is satisfied.

\subsubsection{Constraint on $\beta$}
\label{beta-constraint}

To summarize the previous results, the constraints associated with a
$\Lambda$-CDM-like behavior (i.e., $t_{\Lambda} \lesssim t_0$) and with the
BBN condition (\ref{BBN-phi-1}) can be written as
\beq
\Lambda\mbox{-CDM-like} + \mbox{BBN} \Rightarrow \beta \lesssim
\max\left[1,K_0^{1/(2m)}\right] .
\label{LCDM-BBN}
\eeq

\subsection{Expansion history for models with $K'(\chi_*)=0$ for some $\chi_*>0$}
\label{expansion-history-K0-m}

We now consider the kinetic models (\ref{K-power-2}) and (\ref{K-power-3}),
and more generally kinetic functions such that $K' = 0$ at some positive value $\chi_* >0$.
Then, as $\chi$ rolls down from $+\infty$, following the Klein-Gordon
equation (\ref{KG-int-1}), it will converge at late time to the largest solution $\chi_*$
of $K'(\chi_*)=0$, to obey the asymptotic behavior (\ref{late-time-behavior}).

\subsubsection{Models with $K'<0$ for $\chi>\chi_*$}
\label{Kp-negative}

Let us first discuss the models such as (\ref{K-power-2}), where the derivative $K'$ is negative
at large values of $\chi$, to the right of the largest zero $\chi_*$. In the explicit case
(\ref{K-power-2}), there is a single critical point $\chi_*$, given by
\beqa
\chi_* & = & (-mK_0)^{-1/(m-1)} , \label{chi-star} \\
K_* & = & K(\chi_*) = -1 + \frac{m-1}{m} (-mK_0)^{-1/(m-1)} . \;\;\; \label{K-star}
\eeqa
To recover a cosmological constant behavior at late time, with
$- \bar{p}_{\varphi} \simeq \bar\rho_{\varphi} \simeq \bar\rho_{\Lambda}>0$,
we must have from Eq.(\ref{rho-phi-def}) $-\cM^4 K_* = \rho_{\Lambda}>0$.
This implies $K_*<0$, because $\cM^4>0$, hence
$(-K_0) > (m-1)^{(m-1)}/m^m \sim 1$.
Therefore, scenarios with $(-K_0) \ll 1$ are ruled out.

Then, the remaining scenarios have $\bar{K}'<0$ and $\bar{K}<0$ at all times.
In particular, the effective energy density $\bar\rho_{\varphi}^{\rm eff}$ is now
positive both at late and early times, see Eq.(\ref{w-phi-eff-1}).
Hence, contrary to the class of models (\ref{K-power-1}),
$\bar\rho_{\varphi}^{\rm eff}$ does not need to
change sign and the effective equation of state parameter $w_{\varphi}^{\rm eff}$
does not need to diverge at a time $t_{\rm eff}$.
At late times, we recover a cosmological constant behavior, but instead of $\bar\varphi$
converging to zero as in Sec.~\ref{expansion-history-K0-p}, it is $\dot{\bar\varphi}$ that
converges to a finite value, and the asymptotics (\ref{a-exp-L}) become
\beqa
t \rightarrow \infty & :  &  \bar\rho_{\varphi}^{\rm eff} \simeq - K_* \cM^4 , \;\;
a(t) \sim e^{\sqrt{-K_*/3} \cM^2 t/M_{\rm Pl}} , \nonumber \\
&& \bar\rho \propto a^{-3} , \;\;
\bar\varphi \simeq \sqrt{2 \chi_* {\cM^4}} t . \;\;\;\;\;\;
\label{a-exp-L-neg1}
\eeqa

\paragraph*{$\underline{\mbox{\rm Scenarios with } (-K_0) \gg 1:}\\$}

Scenarios with $(-K_0) \gg 1$ follow the same expansion history
(\ref{I-p})-(\ref{III-p}) as those with $K_0 \gg 1$, and the time $t_{\Lambda}$
that marks the transition between $\bar{K} \simeq K_0 \bar\chi^m$ and
$\bar{K} \simeq K_* \simeq -1$ is given by
 \beq
(-K_0) \gg 1 : \;\; t_{\Lambda} = (-K_0)^{-1/(2m)} \, \beta \, t_0 ,
\label{t-Lambda-mK0-large-1}
\eeq
hence
\beq
\beta \lesssim (-K_0)^{1/(2m)} ,
\label{t-Lambda-mK0-large-2}
\eeq
in a fashion similar to Eq.(\ref{t-Lambda-K0-large}).
Another sign difference is that we now have $\dot{\bar\varphi}>0$ and
$\bar\varphi>0$, from Eq.(\ref{KG-int-1}), because we now have $\bar{K}'<0$.

The time $t_m$, where $\bar{K}'$ shifts from $mK_0\bar\chi^{m-1}$ to $1$ is still
given by Eq.(\ref{tm-def}), where we take $K_0 \rightarrow (-K_0)$.
If $t_m>t_0$, which corresponds to $(-K_0)^{1/[2(m-1)]} \beta > 1$, we still have
the property (\ref{K0-p-betaphi}), with $K_0 \rightarrow (-K_0)$,
\beqa
(-K_0)^{1/[2(m-1)]} \beta > 1 , \;\;  t= t_0 & : & \nonumber \\
&& \hspace{-2.5cm} \frac{\beta\bar\varphi}{M_{\rm Pl}} \sim
\left( \frac{\beta^{2m}}{-K_0} \right)^{1/(2m-1)} \lesssim 1 .
\label{K0-m-betaphi-1}
\eeqa
In this case, the linear term $\chi$ in Eq.(\ref{K-power-1}) is always subdominant
for $t\leq t_0$ and plays no role.
If $t_m<t_0$, which corresponds to $(-K_0)^{1/[2(m-1)]} \beta < 1$, the scalar field
$\varphi$ grows linearly with time in the interval $t_m<t<t_0$ and we obtain today
\beqa
(-K_0)^{1/[2(m-1)]} \beta < 1 , \;\;  t= t_0 & : & \nonumber \\
&& \hspace{-2.5cm} \frac{\beta\bar\varphi}{M_{\rm Pl}} \sim
(-K_0)^{-1/[2(m-1)]} \beta \ll 1 . \;\;\;
\label{K0-m-betaphi-2}
\eeqa
Thus, in both cases the BBN constraint (\ref{BBN-phi-1}) is satisfied

As for the scenario discussed below Eq.(\ref{K0-p-betaphi-2}), the second case
(\ref{K0-m-betaphi-2}) only gives very small deviations from
a uniform quintessence model as it yields $|\beta\bar\varphi/M_{\rm Pl}| \ll 1$.\\

Scenarios with $(-K_0) \sim 1$ are allowed if we accept relative deviations
from $\Lambda$-CDM of order unity. Their scalings can be read from the expressions
above where we set $(-K_0) \sim 1$.

\subsubsection{Models with $K'>0$ for $\chi>\chi_*$}
\label{Kp-positive}

We now turn to models such as (\ref{K-power-3}), where $K'$ is positive for $\chi>\chi_*$.
For the explicit model (\ref{K-power-3}), which has $K_0= 1/4$ and $m=3$, the
largest zero is
\beq
\chi_* = 2, \;\;\; K_* = -1 . \label{K-star-2}
\eeq
Because $K_*<0$ this also corresponds to a positive cosmological constant at late times.
As $K_0$ is of order unity and positive, the expansion history is similar to the one found
in Sec.~\ref{K0-unity}. That is, we recover the three stages (\ref{I-p})-(\ref{III-p})
and the constraint on $\beta$ is $\beta \lesssim 1$.
Because $K_0$ is positive, as for the models of Sec.~\ref{expansion-history-K0-p}
the effective energy density $\bar\rho_{\varphi}^{\rm eff}$ is negative at early times
and it must change sign at a time $t_{\rm eff}$, where the effective equation of state parameter
$w_{\varphi}^{\rm eff}$ diverges.
On the other hand, because of the zero $\chi_*$, $\varphi$ grows linearly with time as
in Eq.(\ref{a-exp-L-neg1}) at late times, but to negative values as
\beqa
t \rightarrow \infty & :  &  \bar\rho_{\varphi}^{\rm eff} \simeq - K_* \cM^4 , \;\;
a(t) \sim e^{\sqrt{-K_*/3} \cM^2 t/M_{\rm Pl}} , \nonumber \\
&& \bar\rho \propto a^{-3} , \;\;
\bar\varphi \simeq -\sqrt{2 \chi_* {\cM^4}} t . \;\;\;\;\;\;
\label{a-exp-L-pos2}
\eeqa

\section{Numerical results}
\label{Numerical-background-1}

We compare in this section the evolution with redshift of the background cosmological
parameters and of the Hubble expansion rate given by the scalar field models with
the reference $\Lambda$-CDM Universe.
All scenarios have the same background cosmological parameters today, taken from
the Planck observations (except of course for the effective dark energy equation of
state parameter, $w^{\rm eff}_{\varphi}$, which is not exactly equal to $-1$ for
the scalar field models).

\subsection{Models with $K'\neq0$}
\label{Kp-positif}

We first consider the class of models (\ref{K-power-1}), with the expansion history
described in Sec.~\ref{expansion-history-K0-p}.

\subsubsection{Dependence on $K_0$}
\label{Dependence-on-K0}

\begin{figure}
\begin{center}
\epsfxsize=8.5 cm \epsfysize=6. cm {\epsfbox{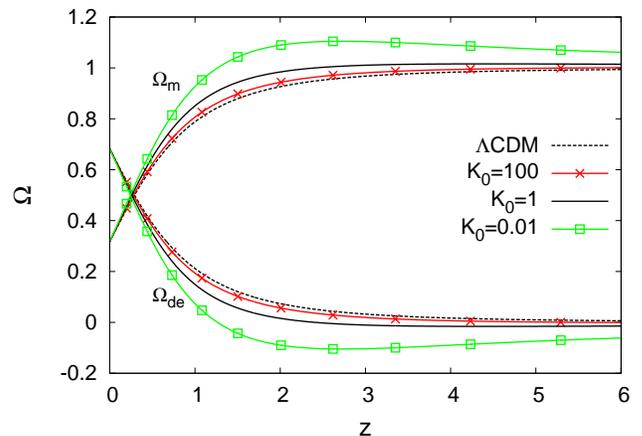}}
\end{center}
\caption{Evolution with redshift of the matter and dark energy cosmological
parameters $\Omega_{\rm m}(z)$ and $\Omega_{\rm de}(z)$. The black dashed lines are
for the reference $\Lambda$-CDM universe and the solid lines are for
the scalar field models (\ref{K-power-1}) with different values of $K_0$
[with $m=3$, $\beta=0.3$, and the exponential coupling (\ref{A-exp-1})].}
\label{fig_Om_z_Kp}
\end{figure}

\begin{figure}
\begin{center}
\epsfxsize=8.5 cm \epsfysize=6. cm {\epsfbox{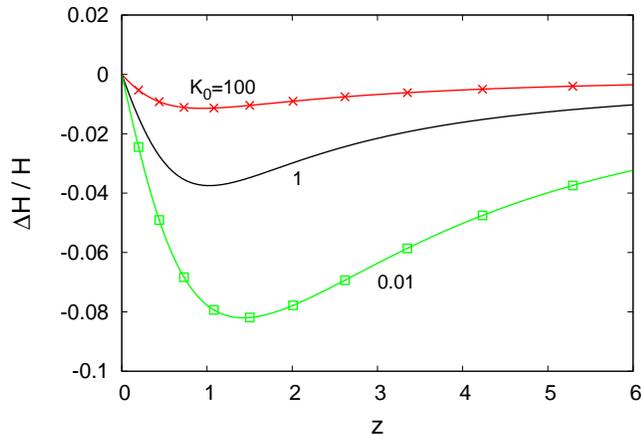}}
\end{center}
\caption{Relative deviation $[H(z)-H_{\Lambda\rm CDM}(z)]/H_{\Lambda\rm CDM}(z)$ of
the Hubble expansion rate with respect to the $\Lambda$-CDM reference.
We consider the same models as in Fig.~\ref{fig_Om_z_Kp}.}
\label{fig_H_z_Kp}
\end{figure}

\begin{figure}
\begin{center}
\epsfxsize=8.5 cm \epsfysize=6. cm {\epsfbox{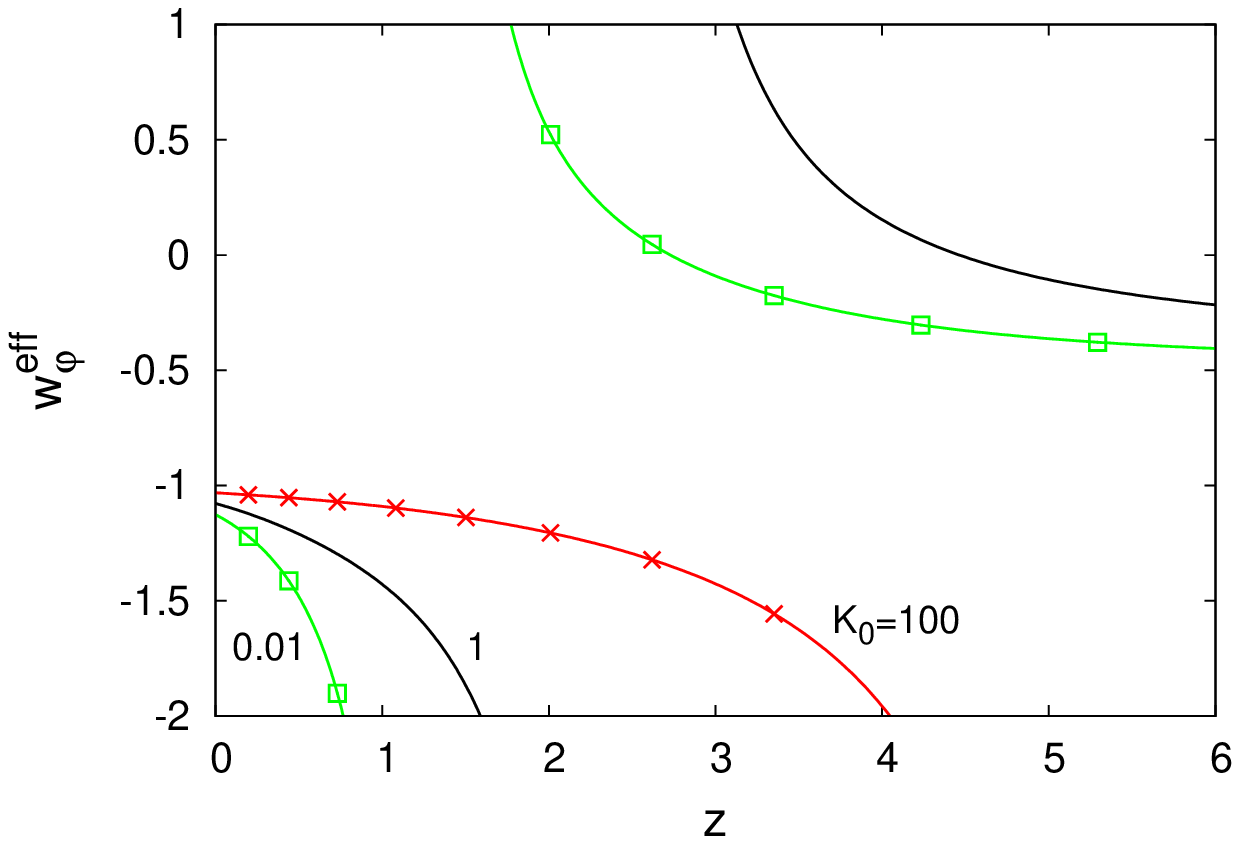}}
\end{center}
\caption{Effective equation of state parameter $w_{\varphi}^{\rm eff}$
for the same models as in Fig.~\ref{fig_Om_z_Kp}.}
\label{fig_w_z_Kp}
\end{figure}

\begin{figure}
\begin{center}
\epsfxsize=8.5 cm \epsfysize=6. cm {\epsfbox{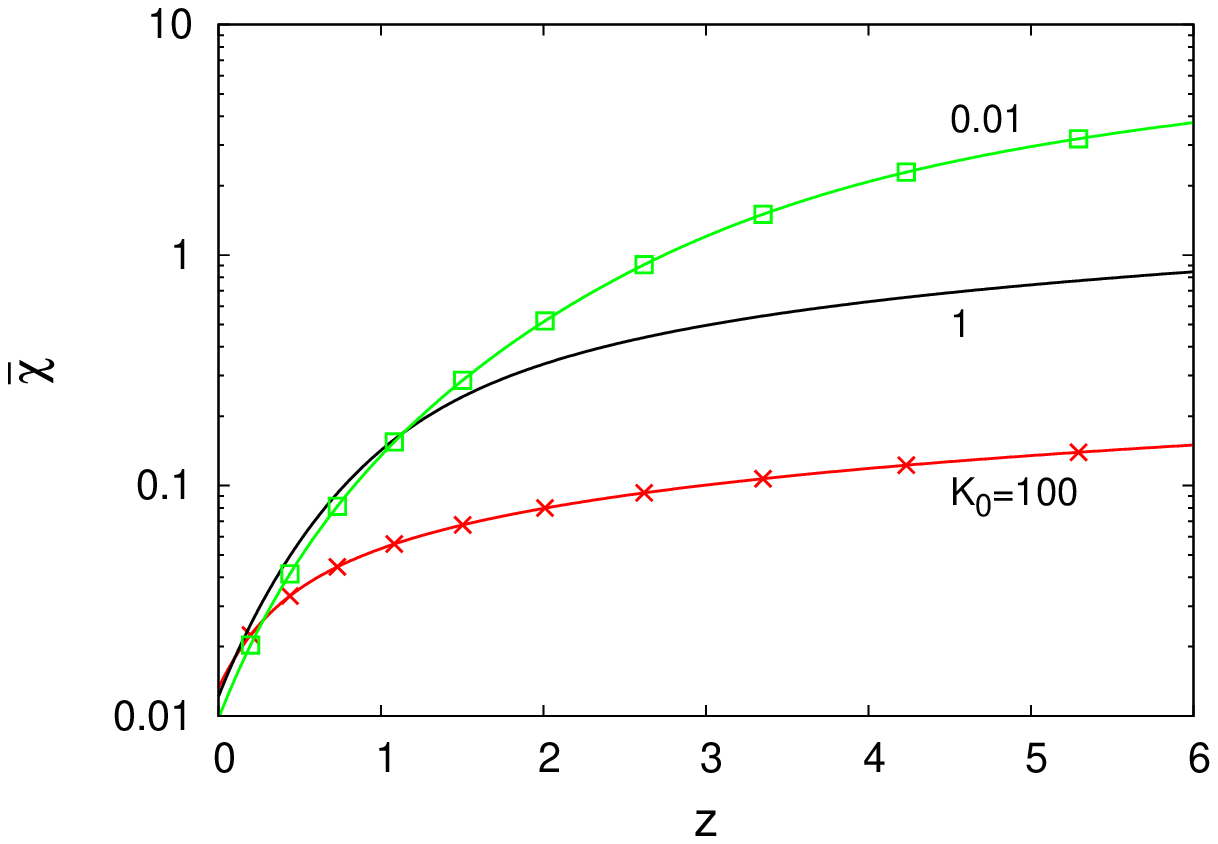}}
\end{center}
\caption{Scalar field time-derivative term $\bar\chi$ of Eq.(\ref{chi-bar-def})
for the same models as in Fig.~\ref{fig_Om_z_Kp}.}
\label{fig_chi_z_Kp}
\end{figure}

\begin{figure}
\begin{center}
\epsfxsize=8.5 cm \epsfysize=6. cm {\epsfbox{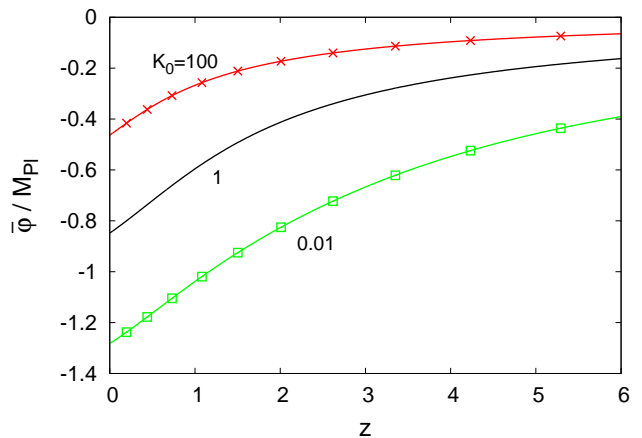}}
\end{center}
\caption{Normalized scalar field $\bar\varphi/M_{\rm Pl}$
for the same models as in Fig.~\ref{fig_Om_z_Kp}.}
\label{fig_v_z_Kp}
\end{figure}

We consider the density parameters $\Omega_{\rm m}(z)$ and
$\Omega_{\rm de}(z)=\Omega^{\rm eff}_{\varphi}(z)$ in Fig.~\ref{fig_Om_z_Kp}.
The four scalar field models have the same coupling function parameter
$\beta=0.3$ and exponential form (\ref{A-exp-1}), and the same cubic exponent
$m=3$ of the kinetic function (\ref{K-power-1}).
They only differ by the value of the kinetic function parameter $K_0$ (and the
derived mass scale $\cM$ set to reproduce the current values $\Omega_{\rm m0}$
and $\Omega_{\rm de0}$).

We consider three positive values, $K_0=100, 1$ and $0.01$, to cover the three
regimes described in Sec.~\ref{expansion-history-K0-p}.
The curves obtained for larger values of $K_0$ are closer to the $\Lambda$-CDM
ones.
This agrees with Eqs.(\ref{t-Lambda-K0-small}) and (\ref{t-Lambda-K0-large}),
which show that $t_{\Lambda}/t_0$ decreases for larger $K_0$
(especially when $K_0>1$). Then, when the scalar field energy density becomes
relevant (at $t_{\rm de} \sim t_0$) the kinetic function $K$ is already very close
to $-1$ and it behaves as a cosmological constant.
As we noticed below Eq.(\ref{a-exp-L}), from Eq.(\ref{w-phi-eff-1}), at high redshift
the effective scalar field density $\bar\rho_{\varphi}^{\rm eff}$ is negative.
This yields a matter density parameter $\Omega_{\rm m}(z)$ that is greater than unity
and a dark energy density parameter $\Omega_{\rm de}(z)$ that is negative
(we always consider a flat universe).

We consider the same set of models in the following.
Thus, we show the Hubble expansion rate in Fig.~\ref{fig_H_z_Kp}.
Again, the behavior is increasingly close to the $\Lambda$-CDM reference
for larger $K_0$, at fixed $\beta$.
Because $\bar\rho_{\varphi}^{\rm eff}$ is negative at high $z$ if $K_0>0$, the Hubble
expansion rate is reduced, in agreement with the Friedmann equation
(\ref{Friedmann-3}).

We display the effective equation of state parameter $w_{\varphi}^{\rm eff}$
of Eq.(\ref{w-def}) in Fig.~\ref{fig_w_z_Kp}.
As explained above, for $K_0>0$ the effective scalar field energy
density $\bar\rho_{\varphi}^{\rm eff}$ is negative at high $z$, which
implies that $w_{\varphi}^{\rm eff}$ diverges at some time $t_{\rm eff}$ and
also changes sign.
In agreement with the previous results and figures, we can check that $t_{\rm eff}$
is pushed farther into the past as $K_0$ increases, so that over the recent period
where $\bar\rho_{\varphi}^{\rm eff}$ is non-negligible we are increasingly close to
$w_{\varphi}^{\rm eff} \simeq -1$ and to the $\Lambda$-CDM behavior.
At high redshift, $w_{\varphi}^{\rm eff}$ converges to $-0.4$ for $m=3$,
see Eq.(\ref{w-phi-eff-1}), independently of the other parameters.
At $z=0$ we have $w_{\varphi}^{\rm eff} \simeq -1$ because we require being close
to the $\Lambda$-CDM cosmology. As explained in Sec.~\ref{expansion-history-K0-p},
this constraint corresponds to an upper bound on $\beta$ or a lower bound
on $K_0$.
In particular, ensuring a value at $z=2$ of $w_{\varphi}^{\rm eff}$
that is sufficiently close to $-1$ requires a large $K_0$ (or a small
$\beta$).

We show the scalar field time-derivative term
$\bar\chi=\dot{\bar\varphi}^2/(2\cM^4)$ in Fig.~\ref{fig_chi_z_Kp}
and the normalized scalar field $\bar\varphi/M_{\rm Pl}$ in Fig.~\ref{fig_v_z_Kp}.
The amplitude of the scalar field and of its time derivative decrease
for larger $K_0$, in agreement with the previous results.
As seen in Sec.~\ref{expansion-history}, the squared time derivative
$\bar\chi$ goes to zero at late time while $\bar\varphi$ converges
to a finite value, see Eq.(\ref{a-exp-L}).
In agreement with Eq.(\ref{K0-sign-phi}), we can check that $\bar\varphi$ and
$K_0$ are of opposite signs.

The fact that $\chi$ grows at early time, even though $\bar\varphi$ goes to zero,
means that the energy density $\bar\rho_{\varphi}^{\rm eff}$ also grows,
as $t^{-2m/(2m-1)}$ as seen in Eq.(\ref{phi-t-0}).
Therefore, the dark energy component becomes subdominant in the past at a slower
rate than a cosmological constant. This leads to the slow decrease with redshift of the
deviation from the $\Lambda$-CDM reference seen in Figs.~\ref{fig_Om_z_Kp}
and \ref{fig_H_z_Kp}.
This means that moderate redshifts, up to $z \lesssim 3$, contain significant information
on the underlying model in this class of scenarios.

\subsubsection{Dependence on $\beta$, $n$, and $m$}
\label{Dependence-on-beta}

\begin{figure}
\begin{center}
\epsfxsize=8.5 cm \epsfysize=6. cm {\epsfbox{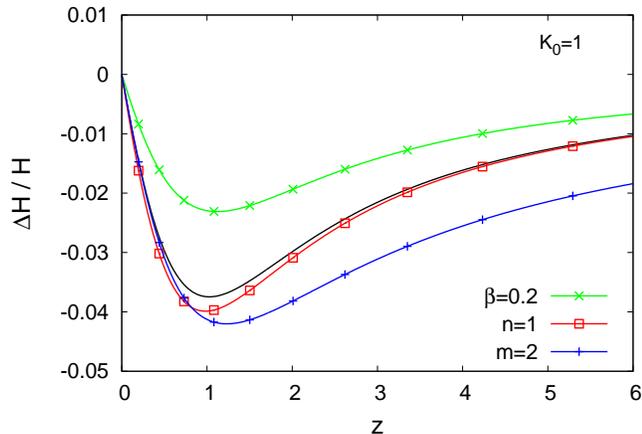}}
\end{center}
\caption{Relative deviation $[H(z)-H_{\Lambda\rm CDM}(z)]/H_{\Lambda\rm CDM}(z)$
of the Hubble expansion rate with respect to the $\Lambda$-CDM reference.
The solid line is the model $\{\beta=0.3,n=\infty;K_0=1,m=3\}$
that was also shown in Fig.~\ref{fig_H_z_Kp}.
The lines with symbols are the results obtained when one of these parameters
other than $K_0$ is modified, either $\beta=0.2$, $n=1$, or $m=2$.}
\label{fig_H_z_K1}
\end{figure}

We show in Fig.~\ref{fig_H_z_K1} the dependence of the Hubble expansion rate
on other parameters of the scalar field model than $K_0$. Taking for reference the case
$\{\beta=0.3,n=\infty;K_0=1,m=3\}$ that was also displayed in
Figs.~\ref{fig_Om_z_Kp}-\ref{fig_v_z_Kp},
we show the results we obtain when we modify one of these other parameters.

Going from the exponential form (\ref{A-exp-1}) (i.e., $n=\infty$) to the
linear form (\ref{A-power-1}) ($n=1$) makes very little change.
This agrees with the fact that we require $|\beta\bar\varphi/M_{\rm Pl}| \lesssim 1$
to comply with the BBN constraint (\ref{BBN-phi-1}), which means that the coupling
function $A(\varphi)$ is well approximated by its first order expansion
$\bar{A} \simeq 1 + \beta\bar\varphi/M_{\rm Pl}$ and the higher order terms, which
depend on $n$, can be neglected when we look for $H(z)$.

Modifying the exponent $m$ of the large-$\chi$ power-law behavior of the kinetic
function $K(\chi)$, from $m=3$ to $m=2$, makes a quantitative change of about
$1\%$ for $H(z)$, but as expected the qualitative shape is not modified.

The main dependence comes from the amplitude $\beta$ of the coupling
function $\bar{A} \simeq 1 + \beta\bar\varphi/M_{\rm Pl}$.
Indeed, at late times $\bar{K}' \simeq 1$ [unless $K_0$ is very large,
as in Eq.(\ref{K0-p-betaphi})], and the main combination that describes the deviation
from $\Lambda$-CDM is $\beta\bar\varphi/M_{\rm Pl} \sim - \beta^2$,
see Eq.(\ref {beta-phi-1}).
Therefore, the amplitude of the deviation from $\Lambda$-CDM is mostly
controlled by $\beta^2$ and it grows with $\beta^2$, as we can check
in Fig.~\ref{fig_H_z_K1}.

\subsection{Models with a fixed point, $K'(\chi_*)=0$}
\label{Kp-zero}

We now consider the classes of models (\ref{K-power-2})-(\ref{K-power-3}), with the expansion
history described in Sec.~\ref{expansion-history-K0-m}.

\begin{figure}
\begin{center}
\epsfxsize=8.5 cm \epsfysize=6. cm {\epsfbox{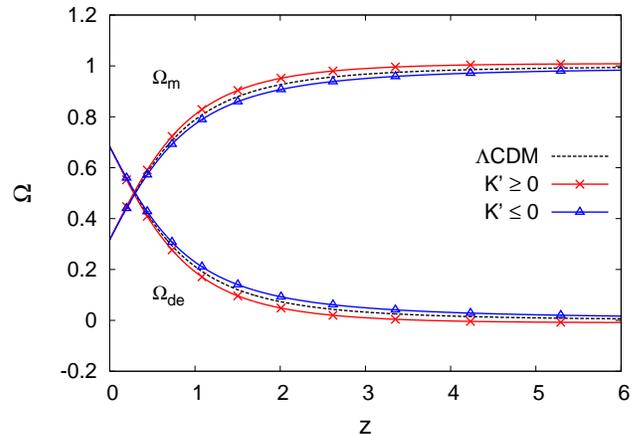}}
\end{center}
\caption{Evolution with redshift of the matter and dark energy cosmological
parameters $\Omega_{\rm m}(z)$ and $\Omega_{\rm de}(z)$. The black dashed lines are
for the reference $\Lambda$-CDM universe and the solid lines with symbols are for
the scalar field models (\ref{K-power-2}), with $\{\beta=0.3,n=\infty;K_0=-5,m=3\}$
(triangles), and (\ref{K-power-3}) (crosses).}
\label{fig_Om_z_Ks}
\end{figure}

\begin{figure}
\begin{center}
\epsfxsize=8.5 cm \epsfysize=6. cm {\epsfbox{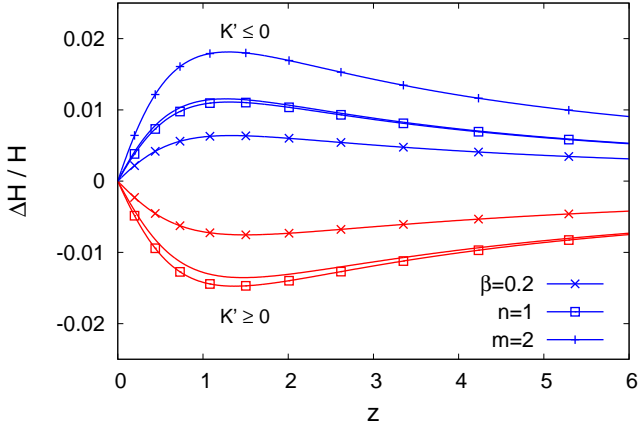}}
\end{center}
\caption{Relative deviation $[H(z)-H_{\Lambda\rm CDM}(z)]/H_{\Lambda\rm CDM}(z)$ of
the Hubble expansion rate with respect to the $\Lambda$-CDM reference.
We consider the same two reference models as in Fig.~\ref{fig_Om_z_Ks}, together with
their variants when we modify either $\beta$, $n$, or $m$ [only for the case of
(\ref{K-power-2})].}
\label{fig_H_z_Ks}
\end{figure}

\begin{figure}
\begin{center}
\epsfxsize=8.5 cm \epsfysize=6. cm {\epsfbox{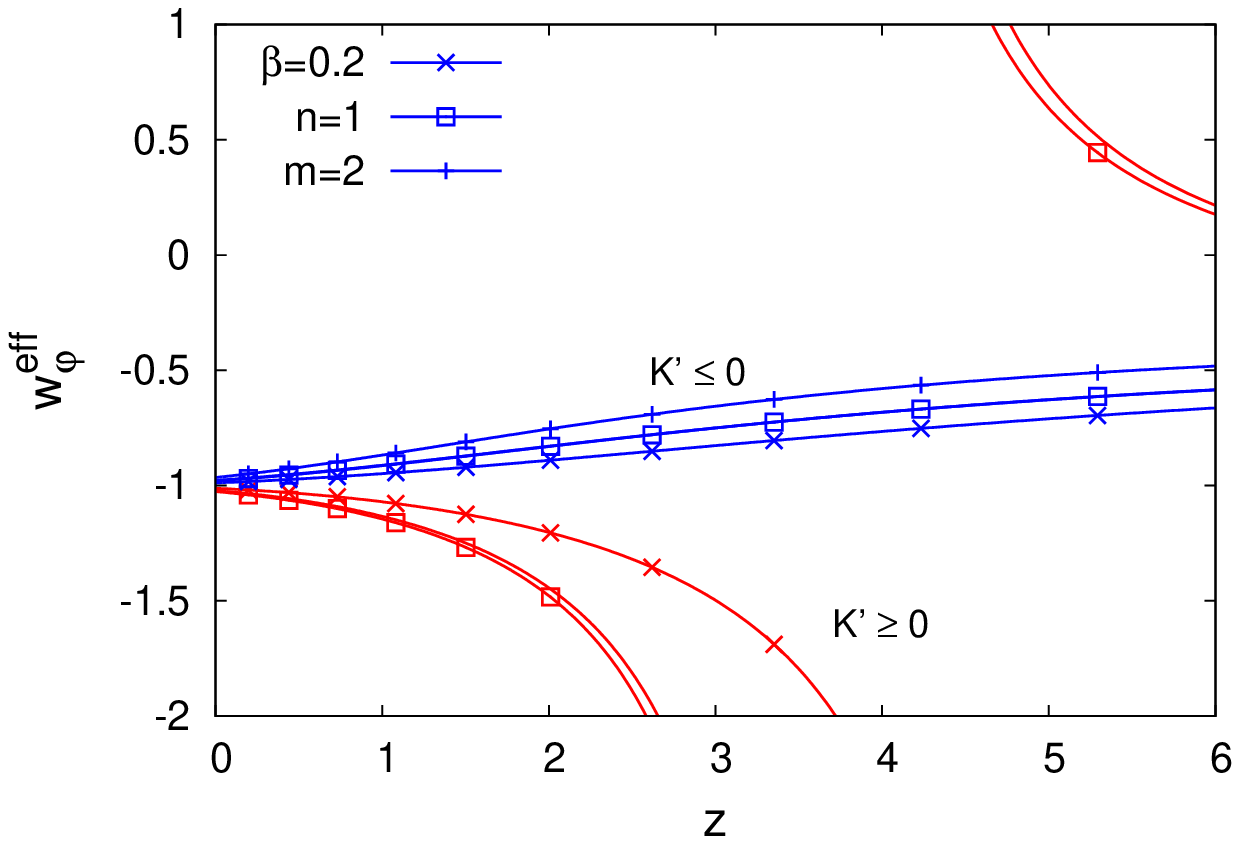}}
\end{center}
\caption{Effective equation of state parameter $w_{\varphi}^{\rm eff}$
for the same models as in Fig.~\ref{fig_H_z_Ks}.}
\label{fig_w_z_Ks}
\end{figure}

\begin{figure}
\begin{center}
\epsfxsize=8.5 cm \epsfysize=6. cm {\epsfbox{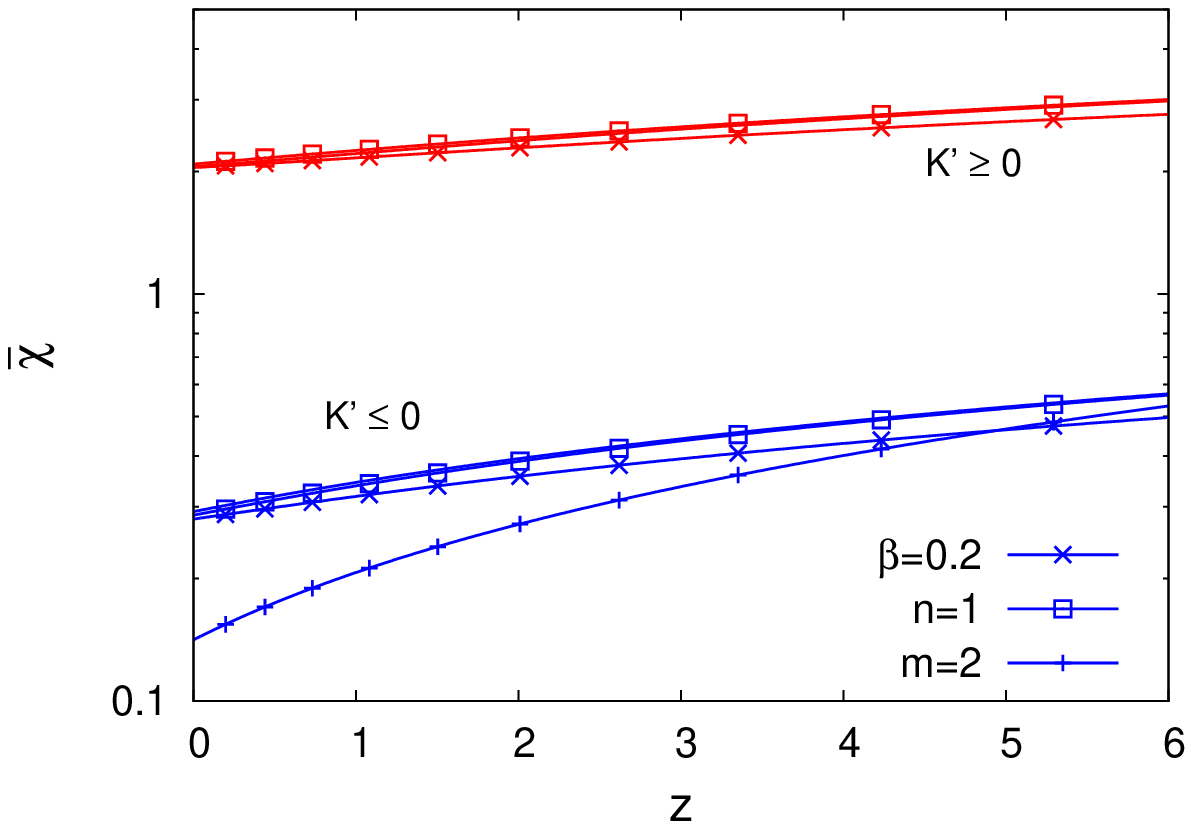}}
\end{center}
\caption{Scalar field time-derivative term $\bar\chi$ of Eq.(\ref{chi-bar-def})
for the same models as in Fig.~\ref{fig_H_z_Ks}.}
\label{fig_chi_z_Ks}
\end{figure}

\begin{figure}
\begin{center}
\epsfxsize=8.5 cm \epsfysize=6. cm {\epsfbox{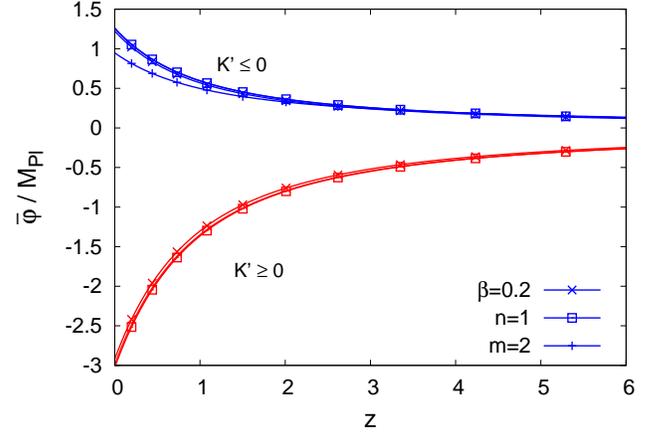}}
\end{center}
\caption{Normalized scalar field $\bar\varphi/M_{\rm Pl}$
for the same models as in Fig.~\ref{fig_H_z_Ks}.}
\label{fig_v_z_Ks}
\end{figure}

We consider the density parameters $\Omega_{\rm m}(z)$ and
$\Omega_{\rm de}(z)$ in Fig.~\ref{fig_Om_z_Ks}.
For the model (\ref{K-power-3}), where $\bar{K}'>0$, we obtain results that are
similar to those obtained in Fig.~\ref{fig_Om_z_Ks}, because at early times it
belongs to the same class with $K_0>0$. Thus, the effective scalar field density
$\bar\rho_{\varphi}^{\rm eff}$ is negative at early times, which yields a matter density
parameter, $\Omega_{\rm m}(z)$, that is greater than unity and a dark energy density
parameter, $\Omega_{\rm de}(z)$, that is negative.

For the model (\ref{K-power-2}), where $\bar{K}'<0$, the effective scalar field density
$\bar\rho_{\varphi}^{\rm eff}$ is always positive, which leads to a matter density parameter,
$\Omega_{\rm m}(z)$, that is smaller than unity and a dark energy density parameter,
$\Omega_{\rm de}(z)$, that is positive, as in the $\Lambda$-CDM scenario.
In fact, in terms of $\Omega_{\rm m}$ and $\Omega_{\rm de}$, the deviation
from $\Lambda$-CDM is of the opposite sign compared to the one
obtained in the case of $K_0 > 0$.

We show the Hubble expansion rate in Fig.~\ref{fig_H_z_Ks}, for these two reference
models and their variants when we modify either $\beta$, $n$, or $m$ [only for the
case of (\ref{K-power-2}) because the model (\ref{K-power-3}) has a fixed $m=3$].
As for the cosmological parameters $\Omega_{\rm m}(z)$ and
$\Omega_{\rm de}(z)$ shown in Fig.~\ref{fig_Om_z_Ks},
the change of sign of $K_0$ corresponds to a change of sign in the deviation
of the Hubble expansion rate from the $\Lambda$-CDM reference. Indeed,
because $\bar\rho_{\varphi}^{\rm eff}$ is negative at high $z$ if $K_0>0$, the Hubble
expansion rate is reduced, in agreement with the Friedmann equation
(\ref{Friedmann-3}). In contrast, if $K_0<0$, $\bar\rho_{\varphi}^{\rm eff}$ is positive
at high $z$ and grows with redshift, see Eq.(\ref{phi-t-0}) (whereas in the
$\Lambda$-CDM scenario $\rho_{\Lambda}$ is constant), so that the
Hubble expansion rate is greater.
Again, the model (\ref{K-power-3}) gives results that are similar to those obtained in
Fig.~\ref{fig_H_z_Kp} for the class (\ref{K-power-1}), because they belong to the same
category at high redshift.

We display the effective equation of state parameter $w_{\varphi}^{\rm eff}$
of Eq.(\ref{w-def}) in Fig.~\ref{fig_w_z_Ks}.
As explained above, for $K_0>0$ the effective scalar field energy
density $\bar\rho_{\varphi}^{\rm eff}$ is negative at high $z$, which
implies that $w_{\varphi}^{\rm eff}$ diverges at some time $t_{\rm eff}$ and
also changes sign. Thus, for the class of models (\ref{K-power-3}) we recover the
behavior found in Fig.~\ref{fig_w_z_Kp}.
For the class of models (\ref{K-power-2}), where $K_0<0$, the effective scalar field
energy density $\bar\rho_{\varphi}^{\rm eff}$ is always positive and $w_{\varphi}^{\rm eff}$
smoothly runs with time from $-(m-1)/(2m-1)$ to $-1$.
This also means that models with $K_0 \lesssim -1$ are very close to
the $\Lambda$-CDM reference, with respect to such background quantities,
because $w_{\varphi}^{\rm eff}$ does not go very far from $-1$, whereas models with
$K_0 >0$ have an equation of state parameter $w_{\varphi}^{\rm eff}$ that goes through
$\pm\infty$ and to ensure a value at $z=2$ that is sufficiently close to $-1$ requires a large
$K_0$ (or a small $\beta$).

We show the scalar field time-derivative term
$\bar\chi=\dot{\bar\varphi}^2/(2\cM^4)$ in Fig.~\ref{fig_chi_z_Ks}
and the normalized scalar field $\bar\varphi/M_{\rm Pl}$ in Fig.~\ref{fig_v_z_Ks}.
Contrary to the models (\ref{K-power-1}) considered in Sec.~\ref{Kp-positif}, at late times
the squared time-derivative $\bar\chi$ goes to a nonzero value $\chi_*$ and
$\bar\varphi$ keeps growing linearly with time, in agreement with
Eqs.(\ref{a-exp-L-neg1}) and (\ref{a-exp-L-pos2}).
The sign of $\bar\varphi$ is opposite to the one of $K_0$ (or more generally $\bar{K}'$).

\subsection{Equation of state $w^{\rm eff}_{\varphi}$}

\begin{figure}
\begin{center}
\epsfxsize=8.5 cm \epsfysize=6. cm {\epsfbox{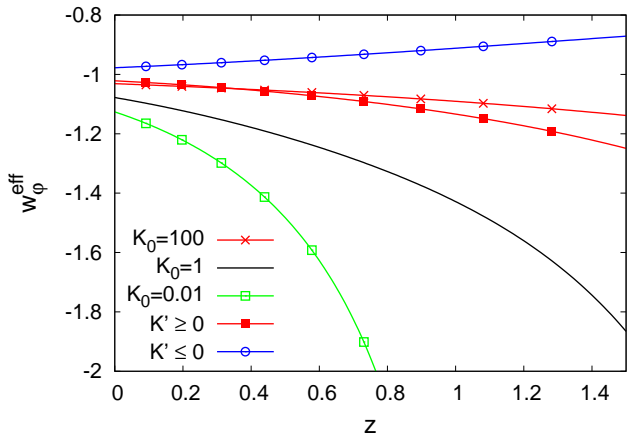}}
\end{center}
\caption{Effective equation of state parameter $w_{\varphi}^{\rm eff}$ for the same models as in Figs.~\ref{fig_Om_z_Kp} and \ref{fig_Om_z_Ks}.}
\label{fig_w_z_zoom}
\end{figure}

We summarize in Fig.~\ref{fig_w_z_zoom} our results for the various models studied in this
paper in terms of the effective equation of state parameter $w^{\rm eff}_{\varphi}(z)$
at low redshift. This shows the variety of behaviors that can be obtained within this
class of K-mouflage scenarios.

\section{Ghosts and K-mouflage}
\label{Ghosts}

The K-mouflage theories that we have considered so far have second order equations of motions. We have seen that their cosmological solutions are classically stable. On the other hand, they can have ghost
instabilities when the kinetic energy becomes negative, as in the class of models
(\ref{K-power-2}) with $K_0<0$.
Because we have specified that the models have a positive sign for the kinetic energy when
$\chi$ is small, see Eq.(\ref{K-chi=0}), this can only happen when $\chi$ is large or beyond
the largest zero $\chi_*$ of $K'(\chi)$, where the dynamics are dominated by the higher order terms in the kinetic
function $K(\chi)$.

We have already seen that the appearance of negative energies for the effective scalar field
energy density $\bar\rho_{\varphi}^{\rm eff}$ in the past implies that the effective equation of state $w_{\varphi}^{\rm eff}$ crosses the phantom divide and even diverges for a redshift which is
$z\gtrsim 1$. This is not a serious problem for the models as the total energy density including matter is always positive. Nonetheless, this puts some constraints on parameters such as $K(0)<0$ for the class of models (\ref{K-power-1}) and $K(\chi_*)<0$
for the classes of models (\ref{K-power-2}) and (\ref{K-power-3}).
The positivity condition, which ensures that $H^2>0$, is automatically satisfied at early times
when $m>1$ because the scalar field becomes subdominant, as described in
Sec.~\ref{early-matter-era}. The condition $m>1$ shows that this behavior can be ascribed to
the effective screening of the K-mouflage field at high cosmological density, because it relies
on the nonlinearity of the kinetic function $K(\chi)$.
Note that the crossing of the phantom divide, associated with $\bar\rho_{\varphi}^{\rm eff}<0$
at early times, actually corresponds to positive scalar field density, $\bar\rho_{\varphi}>0$,
and positive $K_0$, see Eq.(\ref{w-phi-eff-1}).
This change of sign is due to the factor $(\bar{A}-1)\bar\rho$ in Eq.(\ref{rho-phi-eff-def}),
which happens to be of the opposite sign and with a slightly larger amplitude than
$\bar\rho_{\varphi}$ in this regime.

Here we shall investigate the quantum instability of the vacuum due to the existence of states
with negative energies \cite{Cline:2003gs}. This corresponds to the models of the class (\ref{K-power-2}) where
there is no crossing of the phantom divide, $w_{\varphi}^{\rm eff}>-1$.
This can be conveniently analyzed by expanding the Lagrangian around the cosmological
background, $\varphi=\bar\varphi +\delta\varphi$.
This defines an effective field theory for the quantum field $\delta\varphi$,
and introducing
\beq
\delta \chi \equiv \chi - \bar\chi = \frac{1}{2\cM^4} \left[ 2 \dot{\bar\varphi} \frac{\pl\delta\varphi}{\pl t}
+ \left( \frac{\pl\delta\varphi}{\pl t} \right)^2 - \sum_{i=1}^3
\left( \frac{\pl\delta\varphi}{\pl x_i} \right)^2 \right]
\eeq
we obtain from Eq.(\ref {K-def}) the Lagrangian
\beq
{\cal L}_{\delta\varphi} = \cM^4 \sum_{\ell=1}^{\infty} \frac{\bar{K}^{(\ell)}}{\ell !} (\delta\chi)^{\ell} ,
\label{L-dphi-def}
\eeq
where $\bar{K}^{(\ell)}= \frac{\dd^{\ell} K}{\dd\chi^{\ell}}(\bar\chi)$.
If the power $m$ is an integer, or more generally if $K(\chi)$ is a polynomial of the order $m$,
the series (\ref{L-dphi-def}) terminates and it only contains $m$ terms, $\ell \leq m$.
The time dependent cosmological background breaks the Lorentz invariance for the field
$\delta\varphi$, and the lowest order term reads as
\beq
{\cal L}_{\delta\varphi}^{(2)} = \left( \frac{\bar{K}'}{2}+ \bar\chi \bar{K}'' \right)
\left( \frac{\pl\delta\varphi}{\pl t} \right)^2 - \frac{\bar{K}'}{2} \sum_{i=1}^3
\left( \frac{\pl\delta\varphi}{\pl x_i} \right)^2 .
\label{L2-def}
\eeq

For the models (\ref{K-power-1}) and (\ref{K-power-3}), $\bar{K}'$ and $\bar{K}''$ are positive
and the kinetic term (\ref{L2-def}) has the standard sign. There is no ghost but the field
$\delta\varphi$ propagates at a speed, $c_{\delta\varphi}$, that is smaller than the speed of light,
\beq
c_{\delta\varphi}^2 = \frac{\bar{K}'}{\bar{K}'+2\bar\chi \bar{K}''} \; c^2 < c^2 .
\label{c-phi-def}
\eeq

For the models (\ref{K-power-2}), with $K_0<0$ and $m>1$, $\bar{K}'$ and $\bar{K}''$ are negative
and the kinetic term (\ref{L2-def}) has a nonstandard negative sign.
The propagation speed is again smaller than light and given by Eq.(\ref{c-phi-def}), but now
there are ghost instabilities.
Because of the negative sign, the propagator of this theory for the quantum field $\delta\varphi$
propagates negative energy states. These negative energy states destabilize the vacuum as
positive energy particles can be created from nothing, being compensated by the appearance of
ghosts to preserve the conservation of energy.
In this case, the vacuum becomes unstable and the background radiation of those particles created
from the vacuum should be observed. Applying bounds on the spectrum of gamma rays in the
Universe leads to constraints on the highest energies available to the created particles,
i.e. the cutoff energy of the model.

It is convenient to normalize the ghost field as
$\delta\varphi = \phi/\sqrt{-\bar K' - 2 \bar\chi \bar K''}$. The higher order terms in the Lagrangian
are treated as perturbative self-interactions of the scalar field and also involve interactions between
the scalar and the graviton.
They contain terms of the form
\beq
{\cal L}_{\phi}^{(2\ell)} \supset \frac{\bar K^{(\ell)}}
{2^{\ell} \ell! \cM^{4(\ell-1)} |\bar K'+2\chi \bar K''|^{\ell}} [ - (\pl \phi)^2 ]^{\ell} .
\eeq
For the models considered in Sec.~\ref{Specific-models},
where $\bar\chi \sim t^{-2/(2m-1)}$ at early times from Eq.(\ref{phi-t-0}),
this gives terms of the form $[ - (\pl \phi)^2 ]^{\ell} / M^{4(\ell-1)}$, with a time dependent
mass
\beq
M \sim \cM \, (t/t_0)^{-m/[2(2m-1)]} .
\label{M-t-def}
\eeq
This mass grows at early times because of the nonlinearity of the kinetic function $K(\chi)$,
associated with the K-mouflage mechanism. This comes from the increasing kinetic prefactor
in Eq.(\ref{L2-def}) and implies that it is more difficult to create ghosts in the past.
These terms give rise to a linear coupling between the graviton and the ghost of the form
\beq
{\cal L}^{(2\ell)} \supset \frac{h_{\mu\nu}}{M_{\rm Pl}}
\frac{\partial^\mu \phi \partial^\nu \phi (\partial\phi)^{2(\ell-1)}}{M^{4(\ell-1)}} ,
\eeq
(where $h_{\mu\nu} = M_{\rm Pl} \delta g_{\mu\nu}$)
which corresponds to a vertex with $2\ell$ momenta and $2\ell +1$ particles.
Taking into account the vertex
between one graviton and two photons of the type (we only pick one part)
\beq
{\cal L}_\gamma = \frac{h_{\mu\nu}}{M_{\rm Pl}} \partial^\mu A^\rho \partial^\nu A_\rho ,
\eeq
we can draw a vacuum diagram containing the $(2\ell)$ ghosts and two photons created from a
graviton fluctuation from the vacuum. This diagram is obviously kinematically forbidden when the
scalars are not ghosts.
Here on the contrary the two photons with positive energies appear at the same time as
scalars with negative energies. The creation rate is given by
\beqa
&&\Gamma_{(2\ell)} = \int \frac{d^4p}{(2\pi)^4} (\prod_{i=1}^{2\ell} \frac{d^3p_i}{(2\pi)^3 2E_i})
(\prod_{j=1}^2  \frac{d^3p'_j}{(2\pi)^3 2E'_j})\nonumber \\ && \times (2\pi)^4 \delta^{(4)}(p-p'_1-p'_2)
(2\pi)^4 \delta^{(4)}(p-\sum_{i=1}^{2\ell}p_i) \, \vert {\cal M}_{(2\ell)}\vert^2\nonumber \\
\eeqa
where the matrix element is symbolically
\beq
{\cal M}_{(2\ell)} \sim \frac{(\prod_{i=1}^{2\ell} p_i) p'_1 p_2'}
{M^2_{\rm Pl} M^{4(\ell-1)}p^2} ,
\eeq
i.e. one momentum appears for each scalar and each photon, and there is one graviton
propagator. Notice that the dispersion relation for the ghosts is now $E^2=c^2_{\delta\varphi}p^2$.
Cosmologically, Lorentz invariance is broken and we consider the K-mouflage models to be
effective field theories with an explicit cutoff $\Lambda$. This cutoff would correspond to the
highest energies for which the model is defined.
The decay rate of the vacuum becomes then
\beqa
\Gamma_{(2\ell)} & \sim & \left(\frac{\Lambda}{M} \right)^{8(\ell-1)}
\frac{\Lambda^8}{M_{\rm Pl}^4} \\
& \sim & \left( \frac{t}{t_0} \right)^{4m(\ell-1)/(2m-1)} \left(\frac{\Lambda}{\cM} \right)^{8(\ell-1)}
\frac{\Lambda^8}{M_{\rm Pl}^4}  \;\;\;\;\;
\eeqa
which depends on the time when the photons are created through $M(t)$.
Denoting by $n_{(2\ell)}$ the number of photons created by vacuum decays, it satisfies a balance
equation given by
\beq
\frac{d(a^3 n_{(2\ell)})}{dt}= a^3 \Gamma_{(2\ell)} ,
\eeq
which specifies how many photons accumulated since the very early Universe.
The production is dominated by late times when $a^3 \Gamma_{(2\ell)}$ grows with time or
decreases more slowly than $1/t$, which is the current case.
Then, the number of photons today is set by the production at low redshifts with
\beq
t= t_0 : \;\;\; n_{(2\ell)} \sim \frac{\Gamma_{(2\ell)}(t_0)}{H_0} \sim
\left(\frac{\Lambda}{\cM} \right)^{8(\ell-1)}
\frac{\Lambda^8}{M_{\rm Pl}^4 H_0} .
\eeq
The cutoff $\Lambda$ should be above the scale $\cM$, which sets the late time cosmological
behavior associated with the dark energy era.
Therefore, the highest production rate is governed by the highest order operator, $\ell=m$.
The production is essentially occurring for $z\lesssim 1$ and is peaked at energies
$E\sim \Lambda$ corresponding to an estimated spectrum
\beq
\frac{dF}{dE}\sim \frac{\Gamma_0}{\Lambda H_0}, \;\;\; \mbox{with} \;\;\; E \sim \Lambda ,
\;\;\; \Gamma_0 = \left(\frac{\Lambda}{\cM} \right)^{8(m-1)}
\frac{\Lambda^8}{M_{\rm Pl}^4}  .
\eeq
This must be compared with the EGRET  spectrum
\be
\frac{dF}{dE} =7.3 \ 10^{-9} \left( \frac{E}{E_0} \right)^{-2.1} \; (\rm{cm^2.s.MeV.st})^{-1}
\label{EGRET}
\ee
for $E\le E_0=451$ MeV \cite{Sreekumar1998}.
Imposing that the vacuum creates fewer photons than the observed ones
for $E\sim \Lambda$, we get a bound on $\cM$ which depends on $m$. Typically we find
$\Lambda \le 1$ keV for $m=2$ and $\Lambda \le 4$ eV for $m=3$, and $\Lambda \to \cM$ as
$m\gg 1$.
Consequently, the theories with ghosts (\ref{K-power-2}) can only describe the very
low energy physics of the late acceleration of the Universe and cannot be considered as
valid descriptions of physics since BBN.
As a result, these theories seem to be very contrived and are less motivated than the
ghost-free K-mouflage models (\ref{K-power-2}), with $K_0>0$, and (\ref{K-power-3}).
Indeed, the kinetic function (\ref{K-power-3}) contains a term $-\chi^2$ with a ghostlike
negative sign.
However, this does not give rise to the ghost instability discussed above because
during the cosmological evolution we have $\bar\chi > \chi_*$, and we can check that
$\bar K'$ and $\bar K''$ are positive, so that the quadratic term
(\ref{L2-def}) has the standard positive sign.

Here we have made a perturbative analysis around the cosmological background $\bar\varphi$.
This is legitimate because we impose the gamma ray constraint (\ref{EGRET})
before the creation of the field fluctuations $\delta\varphi$ overcomes the background.
Indeed, at the redshifts of interest, $z \lesssim 2$, the scalar field energy density is of the order
of the critical density, $\bar\rho_{\varphi} \sim \bar\rho_c$. The nonperturbative regime corresponds
to $|\rho_{\delta\varphi}| \sim \bar\rho_{\varphi}$, which implies $\rho_{\gamma} \sim \bar\rho_c$
because the energy that goes into the generated photons is the opposite of the one that disappears
in the ghosts. The upper bound $\rho_{\gamma} \lesssim \bar\rho_c$ is much looser than the
EGRET constraint; therefore, we first reach the upper bound (\ref{EGRET}), far in the perturbative
regime. This validates our perturbative analysis and the conclusions that theories with ghosts only make sense at very low energy.

\section{Conclusion}

We have presented a cosmological analysis of  K-mouflage models, one of the three types of theories with screening properties for a long range
scalar interaction on cosmological scales. In this paper, we have focused on the background cosmology. K-mouflage models are characterized by
a nonlinear Lagrangian in the kinetic terms of a scalar field. In dense environments, such as the early Universe, the effects of the scalar field are screened and deviations
from the Einstein-de Sitter cosmology characterizing the matter-dominated era become negligible. At late time, the background cosmology for small redshifts can be taken to be close to the one of an accelerated universe with a cosmological constant.
In between these two epochs, the cosmology of K-mouflage models is rich and differs from the ones of chameleonlike and Galileon models, the other two archetypical models with screening properties.
More precisely, in the early Universe, the screening of the scalar field does not imply that the dark sector of the model converges to the $\Lambda$-CDM model. Indeed, the equation of state of the scalar field converges in the far past to a negative constant, which is not equal to $-1$. At late time and for $z\lesssim 1$, the equation of state is not constant and can evolve by a few percent while other constraints such as the absence of disruption for Big Bang Nucleosynthesis are applied. For moderate redshifts, $1\lesssim z\lesssim 5$, the Hubble rate is significantly different from the $\Lambda$-CDM case. Moreover, in the case where no quantum ghosts and therefore no vacuum instability are present, the equation of state always crosses the phantom divide and even diverges for moderate redshifts. This follows from the change of sign of the effective energy density of the scalar field which goes from negative in the distant past to positive in the recent past. The fact that the scalar energy density becomes negative does not jeopardize the soundness of the models, indeed the scalar becomes more and more screened in the past and the total energy density is always positive.

At the background level and for small redshifts, the K-mouflage models could be tested by observations of the time evolution of the equation of state. At the perturbation level, and as shown in a companion paper \cite{Brax:2014ab}, 
the K-mouflage models are such that large-scale structures are still in the linear regime of the scalar sector. Therefore, deviations from $\Lambda$-CDM on the growth of large-scale structure and on the Integrated Sachs Wolfe effects are present. K-mouflage models are also very different from models like Galileons in the small-scale and large-density regime. The study of the behavior of K-mouflage models in this nonlinear regime is left for future work.

\begin{acknowledgments}

This work is supported in part by the French Agence Nationale de la Recherche under Grant ANR-12-BS05-0002. Ph. B. acknowledges partial support from the  European Union FP7  ITN
INVISIBLES (Marie Curie Actions, PITN- GA-2011- 289442) and from the Agence Nationale de la Recherche under contract ANR 2010 BLANC 0413 01.

\end{acknowledgments}

\bibliography{ref1}   

\end{document}